\documentclass[twocolumn]{aastex631}

\usepackage{amsmath}
\usepackage{amsfonts} 
\usepackage{graphicx,color}
\usepackage{soul}

\begin{document}

\title{A Fast second-order solver for stiff multifluid dust and gas hydrodynamics}

\correspondingauthor{Leonardo Krapp}
\email{krapp@arizona.edu}

\author[0000-0001-7671-9992]{Leonardo Krapp}
\affiliation{Department of Astronomy and Steward Observatory, University of Arizona, Tucson, AZ 85721, USA}

\author[0000-0002-7056-3226]{Juan Garrido-Deutelmoser}
\affiliation{Instituto de Astrofísica, Pontificia Universidad Cat\'olica de Chile, Av. Vicuña Mackenna 4860, 782-0436 Macul, Santiago, Chile}

\author[0000-0002-3728-3329]{Pablo Ben\'itez-Llambay}
\affiliation{Facultad de Ingenier\'ia y Ciencias, Universidad Adolfo Ib\'añez, Av. Diagonal las Torres 2640, Peñalol\'en, Chile}
\affiliation{Data Observatory Foundation, ANID Technology Center No. DO210001}

\author[0000-0001-5253-1338]{Kaitlin M. Kratter}
\affiliation{Department of Astronomy and Steward Observatory, University of Arizona, Tucson, AZ 85721, USA}

% Abstract of the paper
\begin{abstract}
We present MDIRK: a Multifluid second-order Diagonally-Implicit  Runge-Kutta method to study momentum transfer between gas and an arbitrary number ($N$) of dust species. 
The method integrates the equations of hydrodynamics with an Implicit Explicit (IMEX) scheme and solves the stiff source term in the momentum equation with a diagonally-implicit asymptotically stable Runge-Kutta method (DIRK). In particular, DIRK admits a simple analytical solution that can be evaluated with $\mathcal{O}(N)$ operations, instead of standard matrix inversion, which is $\mathcal{O}(N)^3$ . 
Therefore the analytical solution significantly reduces the computational cost of the multifluid method, making it suitable for studying the dynamics of systems with particle-size distributions.
We demonstrate that the method conserves momentum to machine precision and converges to the correct equilibrium solution with constant external acceleration. 
To validate our numerical method we present a series of simple hydrodynamic tests, including damping of sound waves, dusty shocks, a multi-fluid dusty Jeans instability, and a steady-state gas-dust drift calculation.  
The simplicity of MDIRK lays the groundwork to build fast high-order asymptotically stable multifluid methods. 
\end{abstract}


\section{Introduction}

Theories of star and planet formation are grounded in the study of astrophysical fluids. Consequently, many decades of research have been devoted to the development of algorithms to capture their non-linear evolution \citep{LeVequebook,Toro_2009}. For a subset of systems, however, it is insufficient to consider only the evolution of a single species. For example, in cold systems, dust is often the dominant opacity source \citep{Pollack1994,Draine2003,Birnstiel2018}, and thus dictates the thermal evolution of the gas. However, for a range of gas densities and dust particle sizes, the coupling between molecular gas and dust is imperfect. 
The density ratio between gas and different solids may not be spatially or temporally constant,  leading to non-negligible relative velocities, and thus momentum exchange \citep{Epstein1924, Whipple1972, Adachi1976}. 
These multi-species systems are susceptible to a wide range of  multi-fluid instabilites that can substantially alter the evolutionary pathway of a system \citep{Youdin2005,Squire2018,Krapp2020,Hopkins2020}. 
A corollary of the fact that dust controls the thermodynamic behavior of these systems is that it also dramatically impacts the observational signatures of these objects, especially in continuum emission at millimeter - infrared wavelengths, and scattered light at optical wavelengths. Thus accurately predicting the non-uniform distribution of dust grains of various sizes is also vital for predicting and interpreting data from ALMA \citep[e.g.,][]{Andrews2018}, SPHERE \citep[e.g.,][]{Avenhaus2018} and forecoming JWST observations.

To fully capture the impact of dust on astrophysical systems, we require numerical methods that can tackle the problem of momentum exchange between $N$ species in an efficient manner.
Several approaches have been developed to study the interaction between dust and gas numerically. 
So-called ``Hybrid" integration schemes utilize Eulerian grids for the gas, but separate Lagrangian super particles for dust \citep[e.g.,][]{Youdin2007,Balsara2009,Bai2010,MINIATI2010,Yang2016,Mignone2019}. In smoothed particle hydrodynamics (SPH), both gas and dust are treated as Lagrangian particles, with careful choice of the kernel for the drag terms \citep[e.g.,][]{Laibe2012,Laibe2014,Booth2015,Price2018,Stoyanovskaya2018}.
Staggered Semi-analytic Methods \citep[e.g.,][]{Fung2019}, are best applied to tightly coupled distributions, where a staggered time step allows for accurate calculation of dust grain terminal velocity. Finally, there exists a range of Euler-Euler methods where all species utilize the same grids, \citep[e.g.,][]{Johansen2004, Paardekooper2006, Hanasz2010} including first-order multifluid \citep{BenitezLlambay2019}, second-order multifluid \citep{Huang2022}, and single-fluid with modified energy equation \citep{Chen2018}.

Solving for the momentum exchange between gas and particles requires special focus since it introduces source terms (and time scales) that modify the stiffness ratio to the system of equations of motion. 
These source terms are usually modeled as drag forces that depend on the relative velocity between gas and dust (in both the linear and non-linear regime), the dust-to-gas density ratio, and the characteristic stopping time \citep{Whipple1972, Adachi1976}.
In the limit of strong coupling (small stopping time relative to a characteristic time scale) and / or strong feedback (large dust-to-gas density ratio) the time-explicit numerical integration of drag forces requires an excessively small time step in order to preserve stability. Such small timesteps dramatically increase the computational cost of a given calculation.

Alternative methods that circumvent the reduction of the time step are time-implicit methods. 
However it is not straightforward to couple a time-implicit drag-force scheme with a time-explicit numerical method for hydrodynamics, (see e.g. \citealt{Huang2022}). In general, this coupling requires a splitting of the hydrodynamics equations which, if not done properly, downgrades the precision of the numerical solution. Moreover, combined time-implicit and time-explicit methods must still convergence to the correct equilibrium solution.
One example of a time-implicit method was presented in \cite{BenitezLlambay2019}. The method coupled a first-order time-implicit scheme for the drag force with an explicit first-order hydrodynamics solver. This multi-species numerical solver is unconditionally and asymptotically stable and conserves momentum to machine precision. Moreover, the complexity of the scheme is order $\mathcal{O}(N)$, with $N$ the number of fluids \citep{Krapp2020}.

In this work, we present MDIRK: a novel
multifluid second-order method that is asymptotically stable, conserves momentum to machine precision, and has complexity $\mathcal{O}(N)$.
We solve the hydrodynamics equations a two-stage Runge-Kutta Implicit-Explicit (IMEX) method.
In addition, we highlight the benefits of a stiff-accurate multifluid solver, and lay the groundwork to develop even high-order implicit schemes that retain complexity $\mathcal{O}(N)$.

\medskip

This work is organized as follows. In Section\,\ref{sec:numerical_method_intro} we introduce the fluid equations for a multifluid mixture of gas and dust and showcase the numerical method. In Section\,\ref{sec:tests} we present the results from a test-suite and validate the properties of our second-order method. In Section\,\ref{sec:conclusions} we summarize the main results of our work and suggest possible extensions to achieve higher order accuracy.

\section{Numerical Method}\label{sec:numerical_method_intro}

We now describe the numerical method for a 1D system. All of its important properties such as momentum conservation, convergence to the correct equilibrium, and order of accuracy are independent of the number of dimensions. Extensions to 2D and 3D are straightforward, because the drag force does not couple momentum in orthogonal directions.

Consider a system of $N-1$ dust species that exchange momentum with an inviscid gas through drag forces. Each dust species is characterized by a density, velocity, and a characteristic stopping time\footnote{The stopping time encapsulates properties of the given dust species such as grain size, solid density, and composition}, $t_{\rm s}$, or equivalent a rate of momentum exchange with the gas, $\alpha_{\rm d}$.
Gas density and velocity are defined as $\rho_{\rm g}$ and $v_{\rm g}$, respectively. Equivalently, for the $i$-th dust species we define the density and velocity as $\rho_{{\rm d},i}$ and ${v}_{{\rm d},i}$.
We neglect the momentum exchange between dust species. 
We assume an isothermal equation of state and the gas pressure is defined as $P = c^2_{\rm s} \rho_{\rm g}$, with $c_{\rm s}$ as the sound speed.
Under these assumptions, the continuity and momentum equations for gas and dust species are
\begin{align}
    \partial_t \rho_{\rm g} + \partial_x \left( \rho_{\rm g} {v}_{\rm g} \right) &= 0 \,, \label{eq:continuity_gas} \\
    \partial_t \rho_{{\rm d},i} + \partial_x  \left( \rho_{{\rm d},i} {v}_{{\rm d},i} \right) &= 0 \,, \label{eq:continuity_dust} \\
    \partial_t (\rho_{\rm g}{v}_{\rm g}) + \partial_x \left( \rho_{\rm g} {v}^2_{\rm g} + P \right)  &= -\sum^{N-1}_{i=1}   \rho_{{\rm g}} \alpha_{{\rm g},i} \left( {v}_{\rm g} - {v}_{{\rm d},i} \right) + G_{{\rm g}} \,, \label{eq:mom_gas} \\
    \partial_t (\rho_{{\rm d},i}{ v}_{{\rm d},i}) + \partial_x \left( \rho_{{\rm d},i} {v}^2_{{\rm d},i}\right) &= - \rho_{{\rm d},i} \alpha_{{\rm d},i}\left({v}_{{\rm d},i} - { v}_{\rm g} \right) + G_{{\rm d},i} \,, \label{eq:mom_dust}
\end{align}
with $i = 1,\dotsc,N-1$.
The right-hand sides of Equations\,\eqref{eq:mom_gas} and \eqref{eq:mom_dust} containt the drag force terms and any external forces $G_{\rm g}$ and  $G_{{\rm d}, i}$, respectively.  
The rates of momentum exchange are $\alpha_{{\rm g},i}$ and $\alpha_{{\rm d},i}$. Momentum conservation implies
\begin{equation}
    \label{eq:drag_coeff}
    \rho_{\rm g}\alpha_{{\rm g},i} = \rho_{{\rm d},i}\alpha_{{\rm d},i}\,,
\end{equation}
therefore $\alpha_{{\rm g},i} = \epsilon_i \alpha_{{\rm d},i}$,    with $\epsilon_i=\rho_{{\rm d},i}/\rho_{\rm g}$. The collision rates may depend on the density, sound speed, and relative velocity between species.
To rewrite the momentum equations in matrix form, we define
\begin{align}
    {\bf u} &= ( \rho_{\rm g} v_{\rm g}, \rho_{{\rm d},1} v_{{\rm d},1}, \dotsc, \rho_{{\rm d},N-1}v_{{\rm d},N-1} )\,, \label{eq:rhovec} \\
    \text{\boldmath$\mathcal{F}$}  &= ( \rho_{\rm g} v^2_{\rm g} + P, \rho_{{\rm d},1} v^2_{{\rm d},1}, \dotsc, \rho_{{\rm d},N-1}v^2_{{\rm d},N-1} )\,, \label{eq:uvec}\\
    {\bf G} &= ( G_{\rm g}, G_{{\rm d},1}, \dotsc, G_{{\rm d},N-1} )\,,
\end{align}
now, Eqs.\,\eqref{eq:mom_gas}-\eqref{eq:mom_dust} can be cast into the  form
\begin{align}
    \label{eq:conservative_full}
    \partial_t {\bf u} + \partial_x \text{\boldmath$\mathcal{F}$} = {\bf M} {\bf u} + {\bf G}\,,
\end{align}
where ${\bf M}$ is the $N \times N$ matrix.
\begin{align}
\label{eq:Mmatrix}
{\bf M} &=  
\begin{pmatrix}
 -\sum^{N-1}_{k = 1} \epsilon_k \alpha_{{\rm d},k} &   \alpha_{{\rm d},1} &   \alpha_{{\rm d},2} & \dotsc &  \alpha_{{\rm d},N-1} \\
 \epsilon_1\alpha_{{\rm d},1} &  -\alpha_{{\rm d},1}  & 0 & \dotsc & 0 \\
\epsilon_2 \alpha_{{\rm d},2} & 0 & -\alpha_{{\rm d},2} & \dotsc & 0 \\
\dotsc  & \dotsc & \dotsc & \dotsc & \dotsc \\
\epsilon_{N-1}\alpha_{{\rm d},N-1} & 0 & 0 & \dotsc & -\alpha_{{\rm d},N-1} \\
\end{pmatrix} \,.
\end{align}

Note that in our case the matrix ${\bf M}$ is defined from the RHS of the momentum equation in terms of conserved variables, not primitives. This is a subtle distinction with the similar coupling matrix ${\bf M_{\rm BKP19}}$, defined in equation (11) of \cite{BenitezLlambay2019} (see Appendix\,\ref{app:Mbkp}). In this work the matrix ${\bf M}= -{\bf R}{\bf M}_{\rm BKP19}{\bf R}^{-1}$, where ${\bf R}$ a diagonal matrix such that $R_{ii} = \rho_i$. The relation between these matrices, their eigenvectors, and eigenvalues will be utilized to explain the properties of the numerical method.
Both matrices are similar to a real symmetric matrix and therefore diagonalizable.  Moreover, if $\lambda_k$ is an eigenvalue of ${\bf M}$, then $-\lambda_k$ is an eigenvalue of ${\bf M}_{\rm BKP19}$. 
Note that the eigenvalues of ${\bf M}_{\rm BKP19}$ are all real and non-negative (there is only one zero eigenvalue) \citep[see][]{BenitezLlambay2019}.
It is also straightforward to show that if ${\bf x}$ is an eigenvector of ${\bf M}_{\rm BKP19}$, then $\tilde{\bf x} = -{\bf R}{\bf x}$ is an eigenvector of the matrix ${\bf M}$. Since ${\bf 1}^{\rm T}=(1,1,\dotsc,1)$ is eigenvector of ${\bf M}_{\rm BKP19}$,  it follows that $\tilde{\bf 1}^{\rm T}=-(\rho_{\rm g},\rho_{{\rm d},1},\dotsc,\rho_{{\rm d},N-1})$ is eigenvector of ${\bf M}$. 

\bigskip

\subsection{Implicit solutions of the drag-force term} 
\label{sec:dragforce}
Before introducing a numerical method to solve Eqs.\,\eqref{eq:continuity_gas}-\eqref{eq:mom_dust}, we will discuss three different alternatives to solve the momentum equation\,\eqref{eq:conservative_full} neglecting all terms, except for the drag force. The three methods correspond to the first-order fully implicit method of \cite{BenitezLlambay2019}, a second-order fully implicit method \citep[see e.g,][]{Huang2022}, and the diagonally implicit Runge-Kutta method (DIRK) utilized in this work.
Our goal is to highlight the major differences between the three methods when solving the momentum equation
\begin{equation}
 \partial_t {\bf u}  = {\bf M} {\bf u}\,.
    \label{eq:dragforce}
\end{equation}
We define the numerical solution at $t=t_{n}+\Delta t$ of Eq.\eqref{eq:drag_coeff} as  ${\bf u}^{n+1}$.
A first-order fully implicit solution can be obtained after solving
\begin{equation}
    \left({\bf I} - \Delta t {\bf M}\right) {\bf u}^{n+1} = {\bf u}^n\,,
    \label{eq:dragforcefirst}
\end{equation}
whereas a second-order fully implicit method can be cast as
\begin{equation}
    \left({\bf I} - \Delta t {\bf M} + \frac{\Delta t^2}{2} {\bf M}^2\right) {\bf u}^{n+1} = {\bf u}^n\,.
    \label{eq:dragforcesecond}
\end{equation}
These two fully implicit methods converge to the correct equilibrium solution and are asymptotically stable, that is, 
\begin{equation}
    \lim_{\Delta t \rightarrow \infty} |{\bf u}^{n+1}-{\bf u}(\Delta t)|=0\,.
\end{equation}
In particular, utilizing the properties of the matrix ${\bf M}$ it can be shown that the second-order method asymptotically converges as $ \lim_{\Delta t \rightarrow \infty} |{\bf u}^{n+1}-{\bf u}(\Delta t)|\sim 1/\Delta t^2$, whereas the first-order method converges as $ \lim_{\Delta t \rightarrow \infty} |{\bf u}^{n+1}-{\bf u}(\Delta t)|\sim 1/\Delta t$.
Therefore, the fully implicit second-order method provides a more accurate solution for all $\Delta t$.

The advantage of the first-order method is that no matrix inversion is required to obtain the solution \citep{Krapp2020}. The matrix inversion can add a complexity $\mathcal{O}(N^3)$ to the problem, making the second-order method $\sim \times N^2$ more expensive than the first-order method.
The extension of the analytical solution of \cite{Krapp2020} to the second order method is challenging since $ \left({\bf I} - \Delta t {\bf M} + \frac{\Delta t^2}{2} {\bf M}^2\right) $ is not a sparse matrix.

Ideally, we would like a numerical solution that takes advantage of the reduced complexity of the first-order method and preserves the accuracy of the fully implicit second-order method. Thus, in this work, we study and test the properties of a second-order diagonally implicit Runge-Kutta (DIRK) method of the form
\begin{eqnarray}
     \left( {\bf I}-\gamma\Delta t {\bf M} \right) {\bf k}_1 &&= {\bf M}  {\bf u}^n \label{eq:dirk_intro_k1} \\
  \left( {\bf I}-\gamma\Delta t {\bf M} \right) {\bf k}_2 &&= {\bf M}  \left({\bf u}^n + (1-\gamma) \Delta t {\bf k}_1\right)\,,
  \label{eq:dirk_intro_k2}   \\
  {\bf u}^{n+1} &&= {\bf u}^{n} + (1-\gamma) \Delta t {\bf k}_1 + \gamma \Delta t {\bf k}_2\,.
  \label{eq:sol_dirk_intro}
\end{eqnarray}
where $\gamma$ is known as the singular value of the DIRK method \citep{Alexander1977}. This two-stage method is second-order accurate (see Appendix\,\ref{sec:ap_order}). Moreover, the solution of each stage requires the inversion of the matrix $\left({\bf I} - \gamma \Delta t {\bf M}\right)$ and therefore an extension of the analytical solution of \citep{Krapp2020} can be utilized to reduce the complexity to order $\mathcal{O}(N)$. We will show the analytical solution of the DIRK method in Section\,\ref{sec:numerical_method_deriv}. 
We also emphasize that coupling the solution\,\eqref{eq:sol_dirk_intro} with other numerical methods for hydrodynamics is not a trivial task, as standard splitting may fail to converge to the correct equilibrium. In Section\,\ref{sec:numerical_method_deriv} we describe the Implicit-Explicit Runge-Kutta method from \cite{ASCHER1997} to solve the full system of momentum equations\,\eqref{eq:mom_gas}-\eqref{eq:mom_dust} utilizing the solution\,\eqref{eq:sol_dirk_intro}. Continuity equations are then included in the MDIRK method described in Section\,\ref{sec:numerical_method}.

\subsubsection{On the value of $\gamma$ and stiff condition}

Alternative versions of the method of \cite{Alexander1977} can be obtained by varying the coefficients of the Runge-Kutta. 
Depending on the coefficients, the value of $\gamma$ is obtained either by the order conditions (e.g., second-order accuracy as $\Delta t\rightarrow 0 $ and/or as $\Delta t\rightarrow \infty$) or by the asymptotic stability condition $\lim_{\Delta t \rightarrow \infty} |{\bf u}^{n+1}-{\bf u}(\Delta t)|\rightarrow 0$. 
Since the form of Eqs.\,\eqref{eq:dirk_intro_k1}-\eqref{eq:sol_dirk_intro} provides an asymptotically stable solution for any value of $\gamma$ (see Appendix\,\ref{sec:ap_conv}), we will use the order conditions to set its value.  

When $\Delta t \rightarrow 0$, the Taylor series of ${\bf u}^{n+1}$ should match the Taylor series of the solution of Eq.\,\eqref{eq:dragforce} up to second-order, which implies $(2-\gamma)\gamma=1/2$, i.e., $\gamma=1 \pm 1/\sqrt{2}$ (see Appendix\,\ref{sec:ap_order} for details). 
With this choice of $\gamma$, the leading term in the difference between the two Taylor series goes as $\gamma^2 (2\gamma -3)$. Therefore, the smallest error is obtained for $\gamma = 1 - 1/\sqrt{2}$. However, we will show in Section\,\ref{sec:athena_test} that the choice of $\gamma = 1 - 1/\sqrt{2}$ can undershot the solution. While $\gamma = 1 + 1/\sqrt{2}$ provides a less accurate (but still second-order) solution, it ensures a monotonic convergence, which is a numerically desirable property.

Instead of utilizing the Taylor series at $\Delta t \rightarrow 0$, one could also set the value of $\gamma$ by matching the series expansion of ${\bf u}^{n+1}$ and ${\bf u}(\Delta t)$ at $\Delta t \rightarrow \infty$. 
It can be shown that ${\bf u}^{n+1}$ asymptotically converges to the correct solution as $\sim (2\gamma-1)/\gamma^2 \lambda^{-1}_k \Delta t^{-1} + (2-3\gamma)/\gamma^3 \lambda^{-2}_k\Delta t^{-2}$, for some $\lambda_k$ (non-zero) eigenvalue of the matrix ${\bf M}$.
In this case, second-order accuracy is obtained with $\gamma=1/2$, otherwise, the method will converge as a first-order method. 

We note that the two order conditions can not be simultaneously satisfied by the same value of $\gamma$. Therefore, there must be a criterion condition that sets the value of $\gamma$ for a given time step $\Delta t$. We define this criterion as, 
\begin{equation}
\label{eq:gamma_cond}
   \gamma = \left\{
    \begin{array}{ccc}
    1/2 & \textrm{if} & \Delta t > {\rm max}(\alpha^{-1}_{{\rm d},i}) \,,\\[-0.5em] \\
    1\pm 1/\sqrt{2} & \textrm{\,\,\, \quad otherwise} \,, \\
    \end{array}
    \right.
\end{equation}
where ${\rm max}(\alpha^{-1}_{{\rm d},i})$ is the largest stopping time among the dust species.
The choice of $\Delta t > {\rm max}(\alpha^{-1}_{{\rm d},i})$ is only approximate since formally one should look into the non-zero eigenvalues of the matrix ${\bf M}$. For instance, the timescale for the condition\,\eqref{eq:gamma_cond} could be better defined in terms of the largest eigenvalue (non-zero and smallest in absolute value)  of the matrix ${\bf M}$, e.g., $\Delta t > | {\rm max}(\lambda_k) |^{-1}$.
However, it can be shown that $| {\rm max}(\lambda_k) |^{-1}$ is bound between the two largest stopping times of the system \citep[see Appendix D][]{BenitezLlambay2019}. 
This justify our choice of $\Delta t > {\rm max}(\alpha^{-1}_{{\rm d},i})$. 
As we describe below, when $\Delta t \sim {\rm max}(\alpha^{-1}_{{\rm d},i})$ the truncation error is in the asymptotic regime. We therefore define the system of Eqs\,\eqref{eq:continuity_gas}-\eqref{eq:mom_dust} as a ``stiff system" when $\Delta t \geq {\rm max}(\alpha^{-1}_{{\rm d},i})$.

\begin{figure}
\includegraphics[]{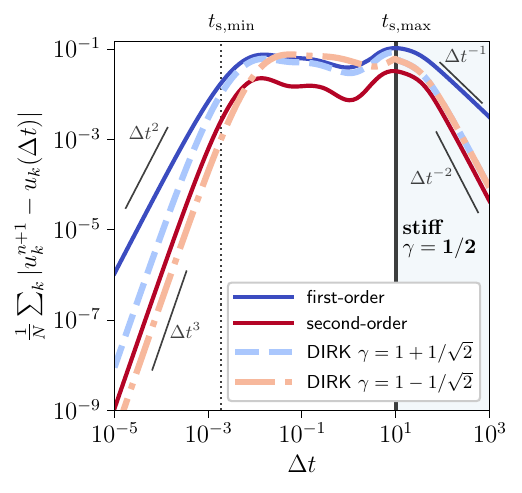}
\caption{Comparison of the (local) truncation error for different methods described in Section\,\ref{sec:dragforce}. The blue solid line corresponds to the first-order implicit method of Eq.\,\eqref{eq:dragforcefirst} whereas the red solid line shows the error for the second-order implicit method of Eq.\,\eqref{eq:dragforcesecond}. The solutions obtained with the DIRK method are shown in blue and red dashed lines.  The black solid line indicates the transition to a stiff regime where $\gamma$ is set to $\gamma=1/2$ for the DIRK method. For $\Delta t \gtrsim 0.002$  (vertical dotted line) time explicit methods will be unstable. 
The analytical solution, $u_k(\Delta t)$, is obtained by solving the eigenvalue problem of the linear equation\,\eqref{eq:dragforce}. }
\label{fig:stiff}
\end{figure}

To highlight the definition of the stiff condition and further compare the methods described in this section, we solve Eq.\,\ref{eq:dragforce} for a system with $N=17$ fluids (16 dust species). 
For each dust fluid, the dust-to-gas mass ratios are obtained as $\epsilon_k = \epsilon (t^{1/2}_{{\rm s}, k+1}-t^{1/2}_{{\rm s}, k})/(t^{1/2}_{\rm s, min}-t^{1/2}_{\rm s, max})$, and $\epsilon=1.0$. 
The stopping time $t_{{\rm s}, k}=\alpha^{-1}_{{\rm d}, k}$ is uniformly spaced in logscale between $t_{\rm s, min}=10^{-3}$ and $t_{\rm s, max}=10$. Note that $t_{\rm s, max} = {\rm max} (\alpha^{-1}_{{\rm d}, i})$ in Eq.\,\eqref{eq:gamma_cond}.
In Figure\,\ref{fig:stiff} we show the truncation error, $e(\Delta t)$, for the three different methods described in this section. The error is obtained as
\begin{equation}
    e(\Delta t) = \frac{1}{N} \sum_k |u_k^{n+1} - u_k(\Delta t)|\,.
\end{equation}
As $\Delta t \rightarrow 0$, the difference between the Taylor series implies that $e(\Delta t) \sim \Delta t^2$ and $e(\Delta t) \sim \Delta t^3$ for the first and second-order methods, respectively.  These scaling are shown in Figure\,\ref{fig:stiff} for $\Delta t \lesssim 0.001$. As already mentioned in this section, the DIRK method is more accurate for $\gamma=1-1/\sqrt{2}$ than $\gamma=1+1/\sqrt{2}$. 
In the regime where $t_{\rm s, min} \lesssim 
\Delta t \lesssim t_{\rm s, max}$, the DIRK method shows an error comparable to that obtained with the first-order implicit method. Similarly, the second-order fully implicit method is also more accurate in this regime. 
As the time-step increases the system reaches the stiff regime where $\Delta t \geq t_{\rm s, max}=10$. In this regime, the numerical solutions show the expected asymptotic convergences $e(\Delta t)\sim \Delta t^{-1}$ and $e(\Delta t)\sim \Delta t^{-2}$ for the first and second-order methods, respectively.
If condition\,\eqref{eq:gamma_cond} is not applied to the DIRK method, i.e. $\gamma=1\pm1/\sqrt{2}$ for all $\Delta t$, the asymptotic convergence will be that of the first-order method. Therefore, without condition\,\eqref{eq:gamma_cond} for $\gamma$  there is not a clear advantage of utilizing a DIRK method in the stiff regime.

\subsection{Second-order Implicit-Explicit Method (IMEX)}\label{sec:numerical_method_deriv}
In this section, we describe the Runge-Kutta IMEX numerical method that will be utilized to solve the hydrodynamics equations.
We begin by focusing on the implementation of the more complex momentum equation; the continuity equation is shown in Section \ref{sec:numerical_method}.
The novelty of this second-order Runge-Kutta scheme, from \citep{ASCHER1997}, is that it incorporates an explicit term and implicit term together in a single time update. As we will demonstrate, the method conserves momentum to machine precision, is asymptotically stable (L-stable), and converges to the correct equilibrium.
We can write  Eq.\,\eqref{eq:conservative_full} as,
\begin{align}
     \partial_t {\bf u} &= \left[ \mathcal{H}({\bf u}) \right]_{\rm exp} +    \left[{\bf M}{\bf u}  \right]_{\rm imp} \label{eq:imex}\,,
\end{align}
 with $\mathcal{H}({\bf u})= - \partial_x  \text{\boldmath$\mathcal{F}$} + {\bf G}$.
The brackets $\left[ \right]_{\rm exp}$ and $\left[  \right]_{\rm imp}$ indicate which terms are time-explicit and time-implicit, respectively.
For our particular system of equations, the term solved with a diagonally implicit method admits an analytical solution \citep{Krapp2020}, which transforms the complexity of the method from $\mathcal{O}(N^3)$ to $\mathcal{O}(N)$, making it suitable for a system with a large number of species. 
Consider the momenta vector ${\bf u}(t_n)={\bf u}^n$. The full update to a new time $t=t_n + \Delta t$ is obtained  as follows:
\begin{align}
   {\bf u}^{n+1} &= {\bf u}^n + \delta \Delta t \mathcal{H}({\bf u}^n) + (1-\delta) \Delta t \mathcal{H}({\bf w}_1) \nonumber \\
   &+ b_1 \Delta t {\bf M}{\bf w}_1 + b_2 \Delta t {\bf M}{\bf w}_2 \label{eq:unp} 
\end{align}
where the coefficients $\tilde{\gamma}$, $\delta$ and $\beta$ are defined in Eq.\,\eqref{eq:coeff}.
The vectors ${\bf w}_1$ and ${\bf w}_2$ are the solution of the following equations:
\begin{align}
   {\bf w}_1 &= {\bf u}^n + \tilde{\gamma} \Delta t \mathcal{H}({\bf u}^n) + \gamma \Delta t {\bf M}{\bf w}_1 \label{eq:w1} \\
   {\bf w}_2 &= {\bf u}^n + \delta \Delta t \mathcal{H}({\bf u}^n) + (1-\delta) \Delta t \mathcal{H}({\bf w}_1) \nonumber \\
   &+ \beta  \Delta t {\bf M}{\bf w}_1 + \gamma \Delta t {\bf M}{\bf w}_2 \,. \label{eq:w2} 
\end{align}
Alternatively, the solution can be also obtained in terms of ${\bf k}_1 = {\bf M}{\bf w}_1$ and ${\bf k}_2 = {\bf M}{\bf w}_2$, by applying the ${\bf M}$ matrix to Eqs\,.\eqref{eq:w1}-\eqref{eq:w2}.
\begin{eqnarray}
     \left( {\bf I}-\gamma\Delta t {\bf M} \right) {\bf k}_1 &&= {\bf M} \left( {\bf u}^n + \tilde{\gamma} \Delta t \mathcal{H}({\bf u}^n) \right) \label{eq:k1} \\
  \left( {\bf I}-\gamma\Delta t {\bf M} \right) {\bf k}_2 &&= {\bf M}  \left({\bf u}^n + \delta \Delta t \mathcal{H}({\bf u}^n) + \beta  \Delta t {\bf k}_1\right) \nonumber \\
  &&+\,\, {\bf M}  \left((1-\delta) \Delta t \mathcal{H}({\bf w}_1)\right)  . \label{eq:k2}   
\end{eqnarray}

The major advantage of the diagonally implicit method is that one first solves for ${\bf k}_1$ and then for ${\bf k}_2$, which simplifies the calculations. Note that this is not the case in fully implicit methods since  Eq.\,\eqref{eq:k1} will also depend on ${\bf k}_2$.
Moreover, we can find the analytical solution of the linear system, which has a complexity $\mathcal{O}(N)$ \citep[c.f.][]{Krapp2020}.
For instance, consider the arbitrary vector ${\bf q}^n = (q^n_{\rm g}, q^n_{{\rm d},1}, \dotsc, q^n_{{\rm d},N-1} )$ and the equation
\begin{equation}
     \left( {\bf I}-\gamma\Delta t {\bf M} \right) {\bf k}^n = {\bf M}{\bf q^n}\,,\label{eq:q}
\end{equation}
the solution ${\bf k}^n = (k^n_{\rm g}, k^n_{{\rm d},1}, \dotsc, k^n_{{\rm d},N-1} )$ has the components
\begin{align}
  {k}^n_{\rm g} &= \frac{ \mathcal{A} - q^n_{\rm g}\mathcal{B} }{1+\gamma\Delta t \mathcal{B}} \label{eq:kg} \\
    {k}^n_{{\rm d},i} &=  \frac{\alpha^n_i}{1+\gamma \Delta t \alpha^n_i} \left( \epsilon^n_i q^n_{\rm g} - q^n_{{\rm d},i} + \gamma \Delta t \epsilon^n_i k^n_{\rm g} \right),  \label{eq:kd}
\end{align}
for $i=1, \dotsc, N-1$, with 
\begin{align}
\label{eq:AB}
    \mathcal{A} &= \sum^{N-1}_{k=1} \frac{\alpha^n_{{\rm d},k} q^n_{{\rm d},k}}{(1+\gamma \Delta t \alpha^n_{{\rm d},k}) } \,, \quad
	\mathcal{B} = \sum^{N-1}_{k=1}\frac{ \epsilon^n_k\alpha^n_{{\rm d},k}}{(1+\gamma \Delta t \alpha^n_{{\rm d},k})}\,. 
\end{align}

It can be shown that $\gamma = 1 \pm 1/\sqrt{2}$ are the only two possible values for $\gamma$ that provide an asymptotically stable and second-order accurate solution (for $\Delta t\rightarrow 0$) with two-stages  \citep{Rosenbrock1963,Alexander1977}.
The rest of the coefficients are chosen to satisfy the order condition and asymptotic behavior of the solution. In this work, we adopt the values from \citep{ASCHER1997} (see also Appendix\,\ref{sec:ap_order})
\begin{align}
\label{eq:coeff}
    \tilde{\gamma} &=\gamma\,,\quad \delta = 1 - 1/(2\gamma)\,, \quad     \beta =1-\gamma\,, \nonumber \\
    b_1 &=1-\gamma\,, \quad b_2=\gamma\,.
\end{align}

An alternative version of the second-order Implicit-Explicit Runge-Kutta can be found in \cite{Pareschi_Russo_2005} with coefficients $\tilde{\gamma}=1$, $\delta = 1/2$, $\beta=1-2\gamma$, $b_1=1/2$ and $b_2=1/2$. 
This alternative version has an explicit part that matches the optimal second-order Runge-Kutta derived in \cite{SHU1988}, with important Total-Variation-Diminishing (TVD) properties. This TVD property also holds for the Runge-Kutta utilized in this work when $\gamma=1+1/\sqrt{2}$, provided a Courant number $C \lesssim 0.58$.

\subsection{MDIRK}\label{sec:numerical_method}

To solve the fluid equations we have reduced the Runge-Kutta described in Section\,\ref{sec:numerical_method_deriv} into a more convenient form, taking into account the continuity equation as well. 
To describe the algorithm we define the discrete vectors at time $t_n$ as ${\bf U}_{\rm g}^n = (\rho^n_{\rm g}, \rho^n_{\rm g} v^n_{\rm g} )= ( U^n_{\rm g 0}, U^n_{\rm g 1} ) $ and  ${\bf K}^n_{\rm g} = (0,k^n_{\rm g})$ for the gas, and ${\bf U}_{{\rm d},i}^n = (\rho^n_{{\rm d},i}, \rho^n_{{\rm d},i} v^n_{{\rm d},i} ) = ( U^n_{\rm d 0}, U^n_{\rm d 1} )$, ${\bf K}^n_{{\rm d},i} = (0,k^n_{{\rm d},i})$ for the $i$-th dust-species, with
$k^n_{\rm g}$ and $k^n_{{\rm d},i}$ defined in Eq.\,\eqref{eq:kg} and Eq.\,\eqref{eq:kd}, respectively. 
Therefore, ${\bf K}_{\rm g}$ and ${\bf K}_{{\rm g},i}$ depend on the density and momentum of all the fluids. 
The zero components of these vectors ensure that the drag force is only considered in the momentum equation. 

The flux of mass and momentum are included in the $\mathcal{L}$ operator, as well as external forces, except of course the drag force. To calculate the fluxes we utilize a standard Riemann solver (see Appendix\,\ref{sec:hydro}). 

\medskip

For each fluid\footnote{Since Eqs.\,\eqref{eq:scheme_full} and \eqref{eq:Up} are agnostic of the type of fluid we omitted the subscript that specifies gas or dust species.}, a full update of the conserved variables from time step $t_n$ to time step $t_{n+1}=t_n + \Delta t$  is obtained as follows 
\begin{align}
\label{eq:scheme_full}
    {\bf U}^* &= {\bf U}^n + \tilde{\gamma} \Delta t \,\mathcal{L}({\bf U}^n)\, \nonumber \\
     {\bf U}^{**} &= {\bf U}^n + (1-\delta) \Delta t \,\mathcal{L}({\bf U}^* + \gamma \Delta t{\bf K}^* )\, \nonumber \\
     {\bf U}^{n+1} &= {\bf U}^p + (b_1-\beta) \Delta t {\bf K}^* + b_2 \Delta t {\bf K}^p \,,
\end{align}
with 
\begin{equation}
\label{eq:Up}
    {\bf U}^p = {\bf U}^{**} + \frac{\delta}{\tilde{\gamma}}( {\bf U}^*-{\bf U}^n) + \beta \Delta t {\bf K}^*\,.
\end{equation}
The coefficients $\tilde{\gamma},\delta, \beta, b_1$ and $b_2$ are given in Eq.\,\eqref{eq:coeff}.
The vectors ${\bf K}^*$ and ${\bf K}^p$ are obtained from the analytical solution given in Eqs.\,\eqref{eq:kg}-\eqref{eq:kd}. When utilizing these analytical solutions, the densities and momenta must be taken from the partial updates ${\bf U}^*$ and ${\bf U}^p$, respectively.
For example,
\begin{eqnarray}
  {\bf K}^*_{\rm g} &&= \begin{bmatrix}
           0 \\
            \frac{\mathcal{A} - U^*_{\rm g1}\mathcal{B} }{ 1+\gamma\Delta t \mathcal{B}} 
         \end{bmatrix}\,, \\ 
         {\bf K}^*_{{\rm d},i} &&= \frac{\alpha^*_i}{1+\gamma \Delta t \alpha^*_i} \left( \begin{bmatrix}
           0 \\
            \epsilon^*_i U^*_{\rm g1} - U^*_{{\rm d1},i}
         \end{bmatrix} +  \gamma \Delta t \epsilon^*_i  {\bf K}^*_{\rm g}\right) 
\end{eqnarray}
with 
\begin{align}
	\mathcal{A} = \sum^{N-1}_{k=1} \frac{ \alpha^*_{{\rm d},k} U^*_{{\rm d 1},k}}{(1+\gamma \Delta t \alpha^*_{{\rm d},k}) } \,, \quad
	\mathcal{B} = \sum^{N-1}_{k=1}\frac{ \epsilon^*_k\alpha^*_{{\rm d},k}}{(1+\gamma \Delta t \alpha^*_{{\rm d},k})}\,.
\end{align}
Note that when the vector ${\bf K}$ is set to zero (no drag forces), second-order accuracy can be obtained by setting $\tilde{\gamma}=1$ and $\delta=1/2$.

\begin{figure*}
\includegraphics[width=\linewidth]{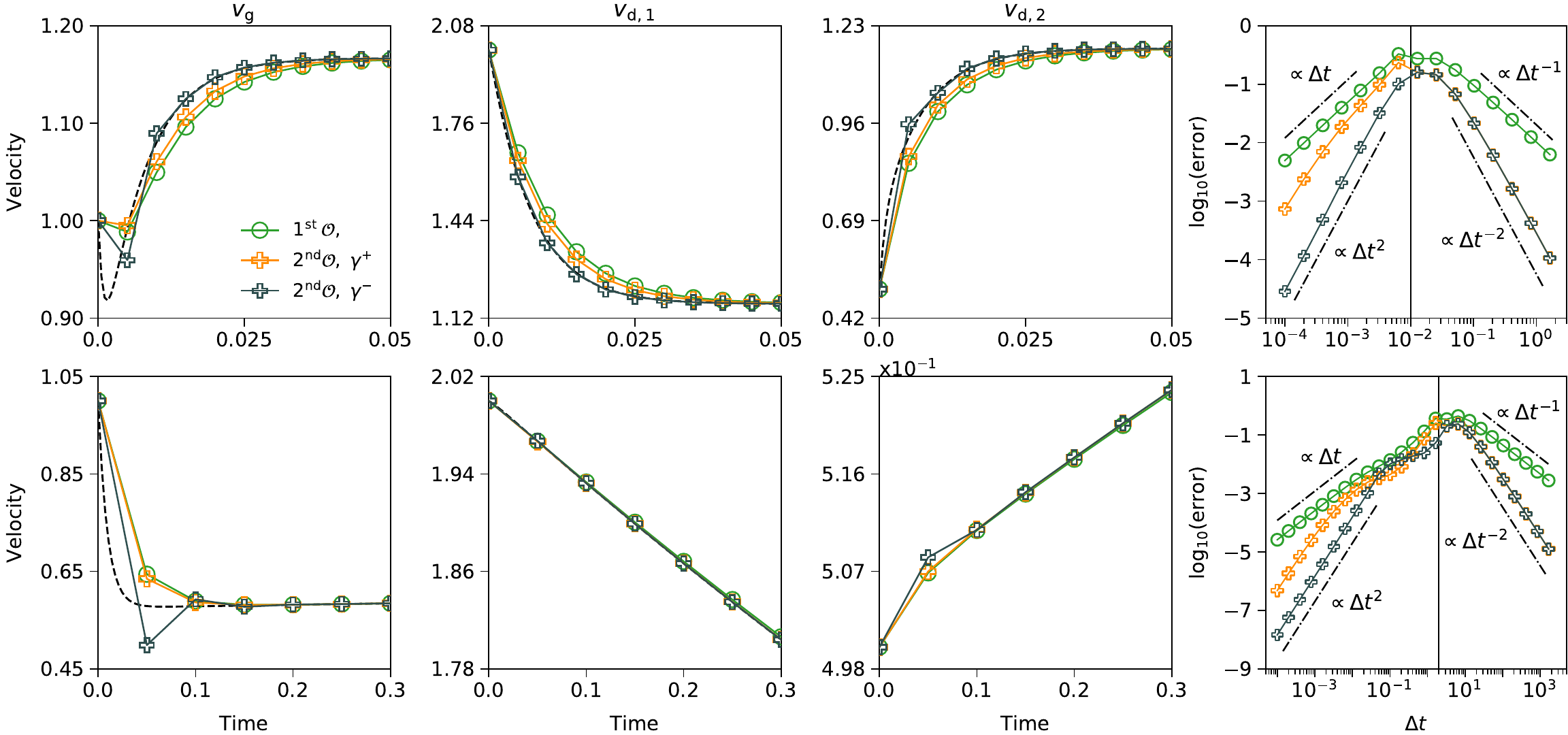}
\caption{From left to right we plot the evolution of the velocity for the gas and two dust species. The dashed line corresponds to the analytical solution, given by Eq\,\eqref{eq:analytic_v}. 
The circles indicate the numerical solutions obtained with different integrators. Green color shows the first-order method, orange and dark lines the second-order method with $\gamma = \gamma^{+} = 1 + 1/\sqrt{2}$ and $\gamma = \gamma^{-} = 1 - 1/\sqrt{2}$, respectively.
The upper and lower panels show the results for the tests 1 and 2, respectively.
In the top and bottom panels, the time step is $\Delta t_{1} = 0.005$, $\Delta t_{2} = 0.05$, respectively (see  Section \ref{sec:athena_test}).
In the rightmost panel, we plot, the error by Eq.\,\eqref{eq:error} as a function of the time step $\Delta t$, for the first-order method (green color) and the DIRK method with $\gamma=1\pm1/\sqrt{2}$. The vertical solid line corresponds to $\Delta t = t_{\rm s, max}$, where condition\,\eqref{eq:gamma_cond} is applied for the DIRK method.   }
\label{fig:stiff_damp}
\end{figure*}

\subsection{Momentum conservation}

The drag-force between species is an internal force that cannot change the total momentum of the system. Our second-order solver has this property as well.  Momentum conservation follows from
\begin{align}
\label{eq:mom}
    U_{{\rm g}1}^{n+1} + \sum_k U_{{\rm d}1, k}^{n+1} &= U_{{\rm g}1}^{n} + \sum_k U_{{\rm d}1, k}^{n} + b_1 \Delta t \left( k^*_{{\rm g}} + \sum_k k^*_{{\rm d},k}\right) \nonumber \\
    &+ b_2 \Delta t \left( k^p_{{\rm g}}+ \sum_k k^p_{{\rm d},k}\right) \nonumber \\
    &= U_{{\rm g}1}^{n} + \sum_k U_{{\rm d}1, k}^{n}\,,
\end{align}
where for any arbitrary ${\bf q}$ (which for example can be taken as ${\bf u}^*$ or ${\bf u}^p$), the vector ${\bf k}$ satisfies 
\begin{align}
     k_{{\rm g}} + \sum_k k_{{\rm d},k} &= k_{{\rm g}} + \sum_k \frac{\alpha_k}{1+\gamma \Delta t \alpha_k} \left( \epsilon_k q_{\rm g} - q_{{\rm d},k} + \gamma \Delta t \epsilon_k k_{{\rm g}} \right) \nonumber \\
   &= k_{{\rm g}} \left(1 + \gamma \Delta t \mathcal{B}\right) - \left(\mathcal{A}-q_{\rm g}\mathcal{B}\right)  \nonumber \\
   &= \frac{\mathcal{A}-q_{\rm g}\mathcal{B}}{(1 + \gamma \Delta t \mathcal{B})}\,\left(1 + \gamma \Delta t \mathcal{B}\right) - \left(\mathcal{A}-q_{\rm g}\mathcal{B}\right) \nonumber \\
   &= 0\,,
\end{align}
recalling $\mathcal{A}$ and $\mathcal{B}$ as defined in Eq.\,\eqref{eq:AB}.  We have also utilized the analytical solutions provided in Eqs.\,\eqref{eq:kg}-\eqref{eq:kd} (see Section\,\ref{sec:numerical_method_deriv}).

\section{Method Test}
\label{sec:tests}
In this section, we present a test-suite that validates our second-order method, and illustrates that the computational costs scales linearly with number of species. We first compare our implicit second-order drag force solver with the first-order implicit solver of \cite{BenitezLlambay2019}. 
Next, we present four test cases using MDIRK method: damped sound waves,  a multi-fluid shock,  an idealized, multi-fluid Jeans instability, and steady-state drift in a 2D shearing box.

\subsection{Damping}\label{sec:athena_test}

As we show in  Appendix \ref{sec:asymptotic}, our drag force solver provides solutions that converge asymptotically to the velocity of the center of mass. 
We show this property and study the accuracy of the method by solving the momentum exchange between gas and two dust species. 
The problem reduces to solving the following equation for the momenta of all fluids
\begin{equation}\label{eq:vector}
    \partial_t {\bf u} = {\bf M} {\bf u} \,.
\end{equation}
for a given initial condition ${\bf u}_0$.
After solving the eigenvalue problem ${\bf M} \bar{\bf x} = \lambda \bar{\bf x}$,  the solution of Eq.\,\eqref{eq:vector} can be expressed as ${\bf u}(t) = \sum_{k=0}^{N-1} \bar{c}_k e^{\lambda_k t}$.
To compare with the work of \cite{Huang2022}, we perform their tests B and C. 
Both tests consider one gas and two dust species with constant stopping times and homogeneous density and velocity backgrounds. The cases differ by a factor of 1000 in damping timescale. In terms of the velocity, the analytical solution corresponds to 
\begin{equation}\label{eq:analytic_v}
    v_{\rm analytical} = v_{\rm CM} + c_1 e^{\lambda_1 t} + c_2 e^{\lambda_2 t}\,,
\end{equation}
where $v_{\rm CM}$ is the velocity of the center of mass. The values of the eigenvalues $\lambda_1$, $\lambda_2$ and the coefficients $c_1$ and $c_2$ can be obtained from table 1 of \citet[][]{Huang2022}. 

In Figure\,\ref{fig:stiff_damp} we show a comparison between the numerical and the analytical solutions, together with the numerical error. 
We include the first-order solution obtained with the method of \cite{BenitezLlambay2019} (green circle curve).
In addition, the relative error with respect to the analytic solution as a function of the time step is displayed in the rightmost panel of Figure\,\ref{fig:stiff_damp}.
The error is obtained as
\begin{equation}\label{eq:error}
    {\rm e}_1 = \frac{1}{M} \sum_{k=0}^{2} \sum_{i=0}^{M-1} \frac{\lvert v_{{\rm numeric},k}(\Delta t_i) - v_{{\rm analytic},k}(\Delta t_i)\rvert}{v_{{\rm analytic},k}(\Delta t_i)} \,,
\end{equation}
where $\Delta t_i$ is the time step and $M$ is the number of time steps during the integration. The index $k=0,1,2$ is the index of gas, and dust species, respectively. 
We calculate the numerical solution for different time steps starting at $\Delta t = 10^{-4}$ until a final time $t_{\rm final} > t_{\rm damp}$, where $t_{\rm damp} \simeq |{\rm max}(\lambda_k)^{-1}|$ is a characteristic damping time to reach the equilibrium velocity. 
When $\Delta t \leq t_{\rm damp}$, the number of times steps $M$ is given by the ratio $t_{\rm damp}/\Delta t$.
When $\Delta t \geq t_{\rm damp}$ we only have one timestep, i.e., $M=1$. 
For the case of test B, $t_{\rm final}=2$ and $t_{\rm damp} \simeq 0.007$, whereas in test C, $t_{\rm final}=2\times 10^3$ and $t_{\rm damp} \simeq 1.9$.

In Figure\,\ref{fig:stiff_damp} we show that the error as a function of $\Delta t$.
In the regime of $\Delta t<t_{\rm damp}$, we obtain  ${\rm e}_1 \sim \Delta t^2$ for both values of $\gamma$ of the DIRK method.
When $\gamma=1+1/\sqrt{2}$, the absolute error is significantly larger in comparison with the solution obtained for $\gamma=1-1/\sqrt{2}$. 
However, for $\gamma=1-1/\sqrt{2}$ the numerical solution may introduce small oscillations about the correct equilibrium.
Fortunately, since the method is asymptotically stable, any initial oscillations are damped to the correct equilibrium. 
This is shown in the left panel of Figure\,\ref{fig:stiff_damp} where the gas velocity initially undershoots the analytical solution  (see Appendix\,\ref{ap:gama} for additional comments on the value of $\gamma$).

When $\Delta t \gg t_{\rm damp}$, the error obtained from Eq.\,\eqref{eq:error} should agree with the characteristic asymptotic convergence rate of the numerical method. 
For large time steps, the error displayed in Figure\,\ref{fig:stiff_damp} converges as $\sim 1/\Delta t$ for the first-order method, whereas for the DIRK method, the error goes as $\sim 1/\Delta t^2$. 
Therefore, the DIRK method provides a second-order accurate solution in the stiff regime. 
We significantly improve the convergence rate of the DIRK method utilizing condition for $\gamma$ given in Eq.\,\eqref{eq:gamma_cond}. The condition sets $\gamma = 1/2$ if $\Delta t \geq t_{\rm s, max}$. In the top panel,  $t_{\rm s, max}=0.01$ whereas in the bottom panel $t_{\rm s, max}=2$. Note that these stopping times are upper bounds for $t_{\rm damp}$ as expected from the discussion in Section\,\ref{sec:numerical_method} and the results on Appendix D of \cite{BenitezLlambay2019}.

 \begin{figure}
\includegraphics[scale=0.5]{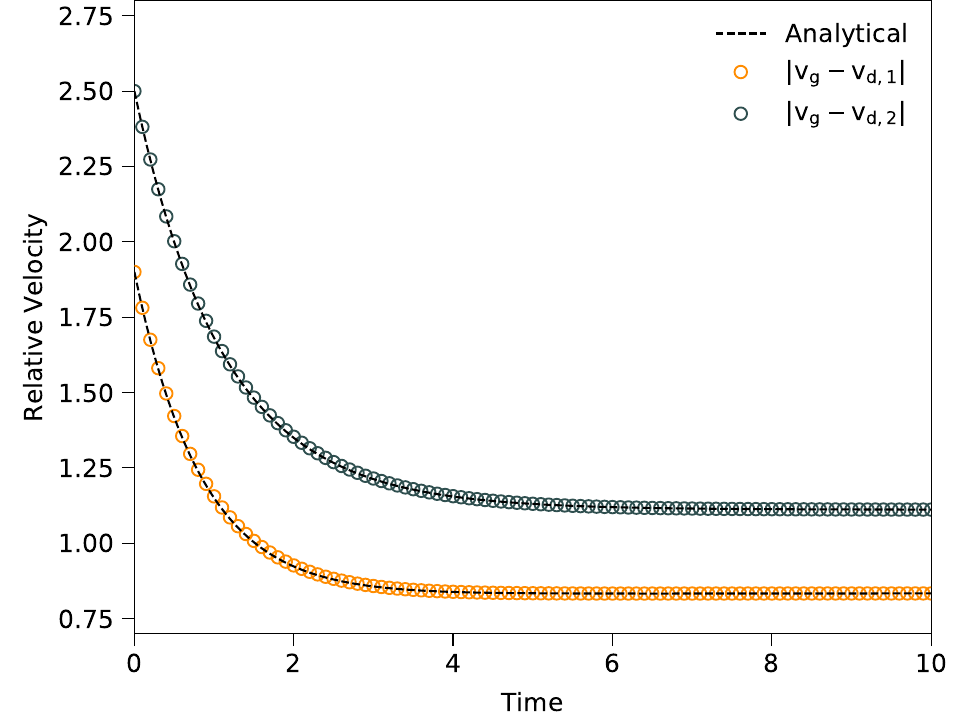}
\caption{Relative velocity between the gas and dust species for the damping test with a constant external force (see Section\,\ref{sec:damping_g}). The numerical solution is shown with orange and blue circles. The dashed line corresponds to the analytical solution. }
\label{fig:constant_acceleration}
\end{figure} 

\subsection{Damping with an external force}
\label{sec:damping_g}

In this section, we extend the test of Section\,\ref{sec:athena_test} by including an external force.  We illustrate the convergence properties of our method when the time-explicit and time-implicit integrators are coupled together as described in Section\,\ref{sec:numerical_method}. 

We again consider one gas and two dust species and solve the momentum equation using Eq.\,\eqref{eq:scheme_full}, neglecting the continuity equation as well as the divergence of fluxes. Under these assumptions, the momentum equation becomes
\begin{equation}
\label{eq:damping_g}
 \partial_t {\bf u} = {\bf M} {\bf u} + {\bf G}\,
\end{equation}
with ${\bf G} = (G_0, 0, 0)$. The analytical solution of Eq.\,\eqref{eq:damping_g} corresponds to,
\begin{equation}
\label{eq:sol_damping_g}
 {\bf u}_{\rm analytical} = {\bf C}\,\,[ t,e^{\lambda_1 t} ,e^{\lambda_2 t}, 1]^{\rm T}\,,
\end{equation}
where $\lambda_1$ and $\lambda_2$ are the non-zero eigenvalues of ${\bf M}$ and ${\bf C}$ is a coefficient matrix that depends on the initial condition, external force, and eigenvalues of ${\bf M}$.
We set as initial condition $\rho_{\rm g} = 1.0$, $\rho_{\rm d1} = 0.1$, $\rho_{\rm d2} = 0.1$ for the density and $v_{\rm g} = 2.0$, $v_{\rm d1} = 0.1$, $v_{\rm d2} = -0.5$ for the velocity. 
The external force is set to $G_0=1$ whereas the drag force coefficients are set to $\alpha_{\rm d,1} = 1.0$, $\alpha_{\rm d,2} = 0.75$, respectively. With these values\, we obtain $\lambda_1 = -1.125$ and $\lambda_2 = -0.8$ and the coefficient matrix ${\bf C}$ is 
\begin{align}
{\bf C} &=  
\begin{pmatrix}
0.833333 & 0.145014 & 0.059615 & 1.79537\\
0.083333 & -0.116011 & 0.029808 & 0.096204\\
0.083333 & -0.029003 & -0.089423 & 0.068426\\
\end{pmatrix} \,.
\end{align}
In Figure\,\ref{fig:constant_acceleration} we show the relative velocity between the gas and the dust species. The dashed line corresponds to the analytical solution, the agreement between the analytical and numerical solutions is
excellent. In the limit of $\Delta t \rightarrow \infty$, the momentum of the $j$-th species (see Eq,\,\eqref{eq:sol_damping_g}) converges to
\begin{align}
\label{eq:sol_damping_g_lim}
    u_j &= P_{j0}\left( \left({\bf P}^{-1}{\bf G}\right)_0\, t +  \left({\bf P}^{-1}{\bf u}\right)_0 \right) - \sum^{N-1}_{k=1} P_{jk} \frac{ \left( {\bf P}^{-1} {\bf G} \right)_k}{\lambda_k}\,
\end{align}
where we assume $\lambda_0 =0$ and $P_{jk}$ denotes the $jk$ entry of the matrix ${\bf P}$, with ${\bf P}$ and ${\bf P}^{-1}$ the matrices such that ${\bf D}={\bf P}^{-1} {\bf M}{\bf P}$ is diagonal, that is, $\lambda_k = {\bf D}_{kk}$. 

After dividing Eq.\,\eqref{eq:sol_damping_g_lim} by the corresponding density we can obtain the relative velocity between species. 
First, note that $P_{j0}/\rho_j - P_{i0}/\rho_i = 0$, for any $i$ and $j$ components. 
This follows from the properties of the matrix ${\bf M}$ described below Eq.\,\eqref{eq:Mmatrix} and using that the first (zero) column of the matrix ${\bf P}$ is the eigenvector $\tilde{\bf 1}^{T}$. 
Therefore, the relative velocity between two species is obtained as
\begin{equation}
   \lim_{\Delta t  \rightarrow \infty} v_j - v_i =  \frac{1}{\rho_i}\sum^{N-1}_{k=1} P_{ik} \frac{ \left( {\bf P}^{-1} {\bf G} \right)_k}{\lambda_k} - \frac{1}{\rho_j}\sum^{N-1}_{k=1} P_{jk} \frac{ \left( {\bf P}^{-1} {\bf G} \right)_k}{\lambda_k}
\end{equation}
In this case we obtain $\lim_{\Delta t \rightarrow \infty} |v_{\rm g} - v_{\rm d1}| = 0.8333 $ and $\lim_{\Delta t \rightarrow \infty} |v_{\rm g} - v_{\rm d2}| = 1.111 $, which are the asymptotic values displayed in Figure\,\ref{fig:constant_acceleration}.
We conclude that the numerical method converges to the correct equilibrium for the case with external constant force. We study this convergence property further in Appendix\,\ref{sec:ap_conv}.

\begin{figure}
\includegraphics[scale=0.5]{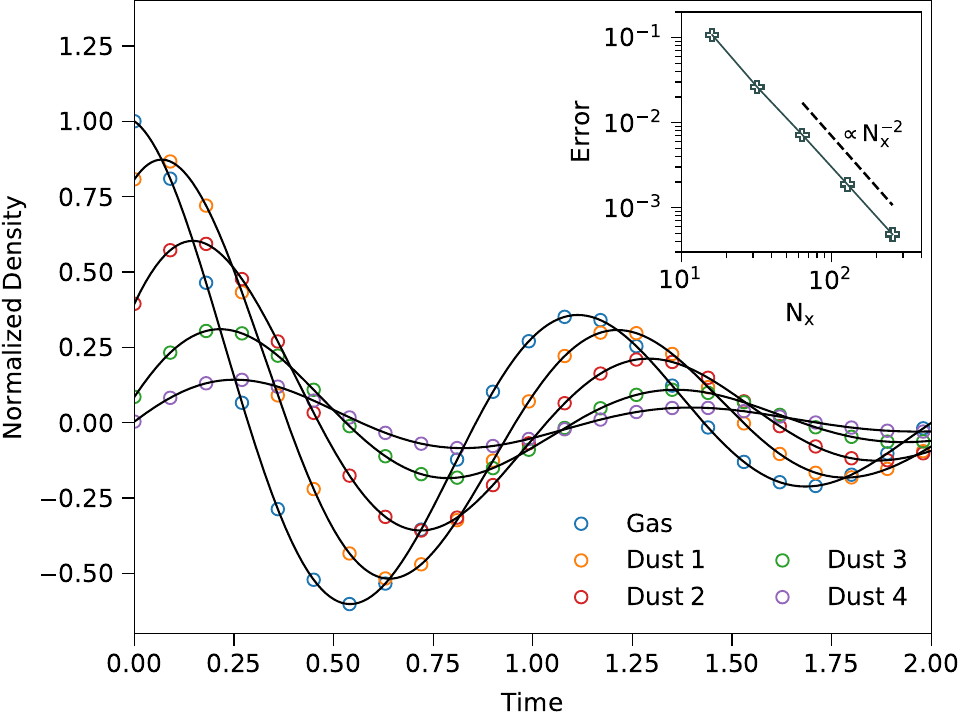}
\caption{Numerical (open circles) and analytical (solid lines) solution of the five-fluid test described in Sec 3.3 in \citep{BenitezLlambay2019}. We plot the time evolution of the normalized density. The blue circles correspond to the gas, while the other color corresponds to the dust species. All solutions were obtained at $x=0$. The normalized density is defined as $\delta \hat{\rho} / (A \rho^0)$, with $A=10^{-4}$. We include in the plot the  $L_1$ error composed of the sum errors of density and velocity (not shown) components of the five fluids. Individual errors are calculated over the whole grid after $t=2.0$.}
\label{fig:damping_soundwave}
\end{figure}
\begin{figure}
\includegraphics[scale=0.5]
{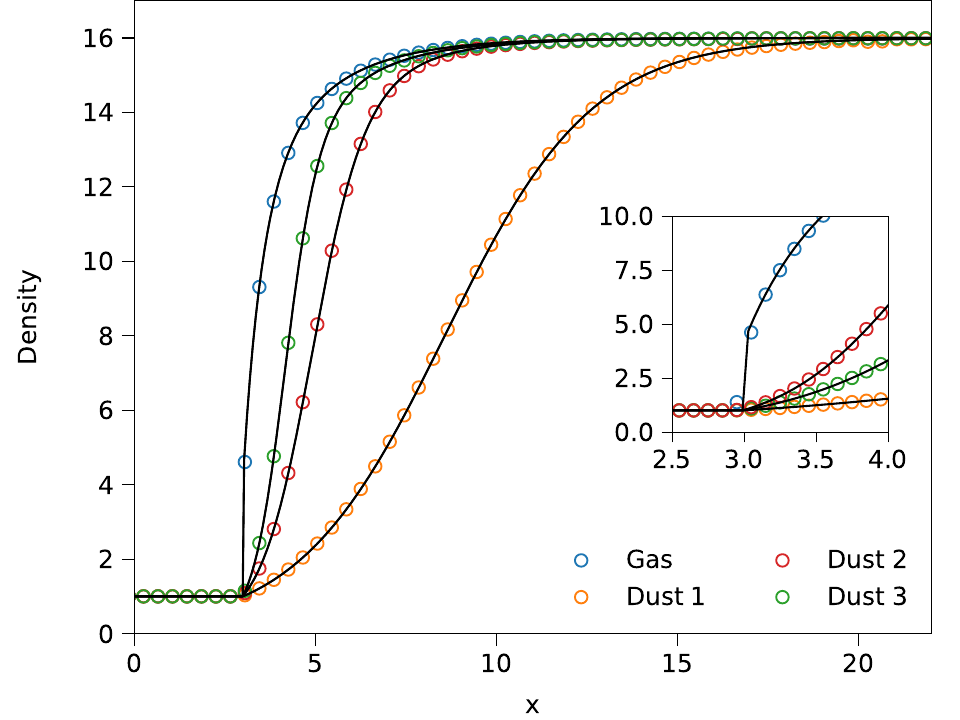}
\caption{Numerical (open circles) and analytical (solid lines) solution of the five-fluid test described in Sec 3.2 in \citep{BenitezLlambay2019}. The numerical solution was obtained at time $t=500$, starting from the initial jump condition. The plot shows the density of the gas (blue) and dust (orange,green,red) species, respectively. In the large panel, to allow the quality of the asymptotic behavior far away from the shock to be assessed, the sampling rate for the open circles was
reduced to 1:4 of the original data. We additionally plot as an inset a zoomed region about the shock showing the full sampling, where the unﬁlled circles correspond to the actual grid points. The code resolves the shock within a single cell, even when an increasing number of ﬂuids is considered. The overall agreement between the numerical and analytical solutions is excellent.}
\label{fig:shock}
\end{figure}  

\subsection{Damping of a soundwave} \label{sec:damping_soundwave}

In this section we reproduce the test described in Section 3.2 of \cite{BenitezLlambay2019}.
For this purpose, we have developed a simple code that solves the Equations\,\eqref{eq:continuity_gas}-\eqref{eq:mom_dust} in a periodic domain with the second-order Runge-Kutta method of Eq.\,\eqref{eq:scheme_full}.
We utilize a uniform mesh in the $x$-coordinate with $N_x$ grid cells.
After obtaining the numerical solution, we compare it with the analytical one and show that our method described in Section\,\ref{sec:numerical_method} achieves the expected second-order accuracy. 

In Figure \ref{fig:damping_soundwave}, we show the analytic (solid lines) and numerical (open circles) solutions for the normalized density, defined as $\delta\hat{\rho}/(\rho A)$.
The test includes five species represented by different colors. Numerical and analytical solutions are in very good agreement.
We study the convergence of the numerical scheme computing the numerical error as
\begin{equation}
    \label{eq:error_l1}    
    {\rm e}_1 = \frac{1}{N_x}\sum_{k=1}^{5} \sum_{i=1}^{2} \left\lvert \delta f_{i,{\rm numeric},k}-\delta f_{i,{\rm analytic},k}\right\rvert \,,
\end{equation}
with $k$ the index of the fluid and $i$ the index of the density and velocity field, respectively.
In the bottom panel of Figure\,\ref{fig:damping_soundwave} we show that the error effectively scales as ${\rm e}_1 \sim N_x^{-2}$. 

\subsection{Shock test}
\label{sec:shock}
In this section we test our numerical method by reproducing the shock test from \cite{Lehmann2018}, extended to multiple fluids in \cite[see Section 3.3][]{BenitezLlambay2019}.
We solve Eqs.\,\eqref{eq:continuity_gas}-\eqref{eq:mom_dust} in an evenly spaced mesh along the $x$-coordinate with 400 grid cells. 
In this shock test, we assume that Eq.\,\eqref{eq:drag_coeff} is constant, that is, $\rho_{\rm g}\alpha_{{\rm g},i}  = \rho_{{\rm d},i}\alpha_{{\rm d},i} \equiv K_i$. Since the drag-force solver has been written in terms of the parameter $\epsilon_i$, the drag-force coefficient $\alpha_{{\rm d}, i}$ must be replaced by $\alpha_{{\rm d}, i}= K_i/\rho_{{\rm d},i}$ in the matrix ${\bf M}$. The details of the test together with initial and boundary conditions are in section 3.3.2 from \cite{BenitezLlambay2019}.

In Figure \ref{fig:shock}, we plot the exact (solid line) and numerical (open circles) solutions for the density of the four fluid shock test. 
To compare the numerical solution with the exact one, we shif the exact solution 2.97 in a $x$-positive direction until the shock position as explained in \citep{BenitezLlambay2019}. The unfilled circles are a sub-sampling (1:4) of the grid points. Inside, we plot a zoomed-in region containing the discontinuity to show the quality of the numerical solution across the shock in the actual grid (all the points are shown).
The Riemann solver captures the shock within one cell. For the dust fluid we utilize the solver described in \cite{Huang2022}.

\subsection{Jeans Instability}\label{sec:jeans}

In order to expand our benchmarking to include codes with self-gravity, we simulate a multifluid Jeans instability in an infinite uniform medium. To our knowledge this is the first implementation of such a test, and so for completeness we present the analytic solution in detail below.

We add the gravitational potential term into Equations (\ref{eq:mom_gas}) and (\ref{eq:mom_dust}), now written  as
\begin{eqnarray}
    \rho_{\rm g} \frac{D{\bf v}_{\rm g}}{D t} &&= \rho_{\rm g} \nabla \Phi - \nabla P - \sum^{N-1}_{i=1} \rho_{{\rm g}} \alpha_{{\rm g},i} \left( {\bf v}_{\rm g} - {\bf v}_{{\rm d},i} \right), \label{eq:jeans1}\\
    \rho_{{\rm d},i} \frac{D{\bf v}_{{\rm d},i}}{D t} &&= \rho_{{\rm d},i} \nabla \Phi - \rho_{{\rm d},i} \alpha_{{\rm d},i}\left({\bf v}_{{\rm d},i} - {\bf v}_{\rm g} \right),
\end{eqnarray}
with $D/Dt=\partial/\partial_t + {\bf v} \cdot \nabla$ the material derivate. The Poisson equation that governs the gravitational potential of the system is
\begin{equation}
    \nabla^{2}\Phi = 4\pi G \left(\rho_{\rm g} + \sum_{i=1}^{N-1} \rho_{{\rm d},i}\right) \,. \label{eq:jeans2}
\end{equation}
In order to study the Jeans instability, we linearize Equations \,\eqref{eq:continuity_gas}-\eqref{eq:continuity_dust}, and \eqref{eq:jeans1}-\eqref{eq:jeans2} with respect to an equilibrium solution obtained after neglecting the contribution of the background gravitational field in the Poisson equation (Jeans Swindle) \citep{Jeans1902,Binney}. 
The equilibrium gas density is constant and denoted as $\rho^0_{\rm g}$. 
The equilibrium dust density of the $i$-th species is also a constant, equal to $\rho^0_{{\rm d},i} = \epsilon_{i}^0 \rho_{\rm g}^0$. Finally, the equilibrium velocities are set to zero. We define small density perturbations $\delta \rho_{\rm g}$ and $\delta \rho_{{\rm d},i}$,
whereas the velocity perturbations are ${\delta \bf v}_{\rm g} = \delta v_{\rm g} \hat{x}$, $\delta {\bf v}_{{\rm d},i} = \delta v_{{\rm d},i} \hat{x}$. We set $\rho^0_{g}=1.0$ and the total dust-to-gas mass ratio $\epsilon^0 = \sum^{N-1}_{i=1} \epsilon^0_i = 0.01$, with $\epsilon^0_i = \epsilon^0/N$. We consider $N=128$ fluids with a uniform distribution of stopping times between $t_{\rm s} = 10^{-4} - 10$

The linearized continuity, momentum, and Poisson equation are 
\begin{eqnarray}
\frac{\partial \delta\rho_{\rm g}}{\partial t} + \rho_{\rm g}^0 \frac{\partial \delta v_{\rm g}}{\partial x} &&= 0 \,, \label{eq:gas_con_pert}\\\\
\frac{\partial \delta\rho_{{\rm d},i}}{\partial t} + \rho_{{\rm d},i}^0 \frac{\partial \delta v_{{\rm d},i}}{\partial x} &&= 0 \,, \label{eq:dust_con_pert}\\
\frac{\partial\delta v_{\rm g}}{\partial t} + \sum_{m=1}^{N-1} \frac{\epsilon_m^0}{t_{{\rm s},m}} (\delta v_{\rm g}-\delta v_{{\rm d},m})+\frac{c_{\rm s}^2}{\rho_g^0}\frac{\partial\delta\rho_{\rm g}}{\partial x} &&= -\frac{\partial\delta\phi}{\partial x} \,, \label{eq:gas_mom_pert}\\
\frac{\partial\delta v_{{\rm d},i}}{\partial t}+\frac{1}{t_{{\rm s},i}}(\delta v_{{\rm d},i} - \delta v_{\rm g}) &&= -\frac{\partial\delta\phi}{\partial x} \,, \label{eq:dust_mom_pert}\\
4\pi G (\delta \rho_{\rm g} + \sum_{m=1}^{N-1} \delta\rho_{{\rm d},m})  &&= \nabla^2 \delta \phi \label{eq:poisson_pert}\,.
\end{eqnarray}
For $N$ fluids, these $2N+1$ equations are coupled, due to the gravitational field and the drag-force.

After assuming plane-wave perturbations $\delta f = \delta\hat{f}e^{i{\bf k}\cdot{\bf x}-\omega t}$ with ${\bf k}$ a real wave-number, we obtain the eigenvalues and eigenvectors of the system.
We normalize the set of linearized equations defining %
\begin{eqnarray}
&&\tilde{\omega} = \omega/(c_{\rm s} k_{\rm J})\,, \quad \quad  \tilde{k}=k/k_{\rm J} \\%
&&\tilde{\rho}_{\rm g} = \delta \rho_{\rm g}/\rho_{\rm g}^0\,, \quad  \quad \tilde{\rho}_{{\rm d},i} = \delta \rho_{{\rm d},i}/\rho_{\rm g}^0 \\
&&\tilde{v}_{\rm g} = \delta v_{\rm g}/c_{\rm s}\,, \quad  \quad \tilde{v}_{{\rm d},i} = \delta v_{{\rm d},i}/c_{\rm s} \\
&&\beta_i = 1/(t_{{\rm s},i} c_{\rm s} k_{\rm J}) 
\end{eqnarray}
where
\begin{align}   
     k_{\rm J} = \sqrt{4\pi G \rho_{\rm g}^0 (1+\epsilon^0)}/c_{\rm s},\, \rm{ and } \quad 
    \epsilon^0 = \sum_{m=1}^{N-1} \epsilon_m^0.
\end{align}

After normalization, \eqref{eq:gas_con_pert}$-$\eqref{eq:poisson_pert} become 
\begin{align}
    -\tilde{\omega} \tilde{\rho}_{\rm g} + i \tilde{k}\tilde{v}_{\rm g} = 0\,, \\
    -\tilde{\omega} \tilde{\rho}_{{\rm d},i} + i \epsilon^{0}_{i} \tilde{k}\tilde{v}_{{\rm d},i} = 0\,, \\
    -\tilde{\omega}\tilde{v}_{\rm g} + \sum_{m=1}^{N-1}\epsilon_{m}^{0}\beta_{m}(\tilde{v}_{\rm g} - \tilde{v}_{{\rm d},i}) + i\tilde{\rho}_{\rm g}\tilde{k} = - i \tilde{k} \frac{\delta\hat{\phi}}{c_{\rm s}^{2}}\,, \\
    -\tilde{\omega}\tilde{v}_{{\rm d},i} + \beta_{i} (\tilde{v}_{{\rm d},i} - \tilde{v}_{\rm g}) = - i\tilde{k} \frac{\delta \hat{\phi}}{c_{\rm s}^{2}}\,, \\
    \frac{\delta \hat{\phi}}{c_{\rm s}^{2}} = -\frac{1}{\tilde{k}^{2}(1+\epsilon^{0})} \left(\tilde{\rho}_{\rm g} + \sum_{m}^{N-1} \tilde{\rho}_{{\rm d},m}\right)\,,
\end{align}
and can be reduced to a $2N$ equation system, written as
\begin{eqnarray}
    i \tilde{k}\tilde{v}_{\rm g} 
    &&= \tilde{\omega} \tilde{\rho}_{\rm g} \,,\\
    i \tilde{k}\epsilon^0_{i}\tilde{v}_{{\rm d},i} 
    &&= \tilde{\omega} \tilde{\rho}_{{\rm d},i} \,, \\
     \tilde{\omega}\tilde{v}_{\rm g} &&= \left(i \tilde{k} - \frac{i}{\tilde{k}(1+\epsilon^0)}\right) \tilde{\rho}_{\rm g} 
    -\frac{i}{\tilde{k}(1+\epsilon^0)} \sum_{m=1}^{N-1} \tilde{\rho}_{{\rm d},m} \nonumber \\
    &&+ \sum_{m=1}^{N-1} \epsilon_{m}^0 \beta_{m} \tilde{v}_{\rm g} 
    - \sum_{m=1}^{N-1} \epsilon_{m}^0 \beta_{m} \tilde{v}_{{\rm d},m}\,\\
    \tilde{\omega}\tilde{v}_{{\rm d},i} &&= - \frac{i}{\tilde{k}(1+\epsilon^0)} \tilde{\rho}_{\rm g}
    - \frac{i}{\tilde{k}(1+\epsilon^0)} \sum_{m=1}^{N-1} \tilde{\rho}_{{\rm d},m} \nonumber \\
    &&- \beta_{i}\tilde{v}_{\rm g} + \beta_{i}\tilde{v}_{{\rm d},i}  \,.
\end{eqnarray}
The equivalent matrix equation is ${\bf A} {\bf V} = \tilde{\omega}{ \bf V}$, ${\bf V} = (\tilde{\rho}_{\rm g},\tilde{\rho}_{{\rm d},i},\dotsc,\tilde{\rho}_{{\rm d},N-1},\tilde{v}_{\rm g},\tilde{v}_{{\rm d},i},\dotsc,\tilde{v}_{{\rm d},N-1})$. The eigenvalues and eigenvectors of the problem, in our tests, have been obtained with the function \textit{eig()} of the \textsc{Numpy} Python package.

\begin{figure}
\includegraphics[width=\columnwidth]{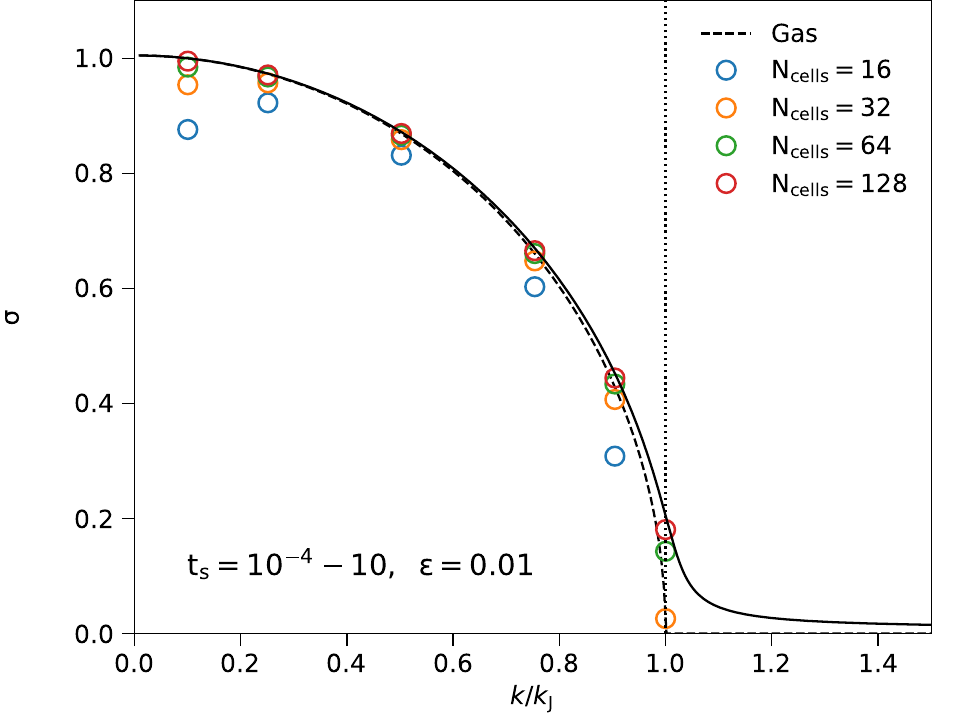}
\caption{Test resolution convergence for Jeans instability including drag forces. The dashed line denotes the analytic solution for only gas. The solid line shows the analytic solution for 128 fluids including gas a dust species with a linear distribution from $t_{\rm s} = 10^{-4}$ to $10$ and uniform density distribution with a total dust-to-gas ratio $\epsilon^{0} = 0.01$. The open circles indicate that the numerical solutions for different numbers of cells converge to the solid line. The vertical dotted line indicates $k/k_{\rm J} = 1/\sqrt{1+\epsilon^{0}}$.}
\label{fig:jeans}
\end{figure}

Figure \ref{fig:jeans} displays the grow rate of the dusty Jeans instability $\sigma = -\min\lbrace\Re(\tilde{\omega})\rbrace\sqrt{1+\epsilon^{0}}$ as a function of $k/k_{\rm J}$. Note that when $\epsilon_0=0$, $\sigma$ reduces to the normalized growth rate of the gas-only problem. Thus, the inclusion of dust enables higher growth rates as shown in Figure\,\ref{fig:jeans} (see solid line vs dashed line (dust-free case).
Comparing the gas only and dusty cases illustrates that while both feature a growth rate that is larger for lower wave numbers, the critical $k/k_{\rm J}$ for stability is shifted to larger values as dust density increases.
In Figure\,\ref{fig:jeans} we also the  numerical solution obtained with MDIRK described in Section\,\ref{sec:numerical_method}. We calculate the growth rate varying the number of cells from $N_{\rm cells} = 16$ to $N_{\rm cells} = 128$. The growth rate converges to the analytic solution for all unstable modes considered in this work\footnote{The largest wavenumber considered in this work corresponds to the cutoff mode of the dust-free case, i.e.,   $k/k_{\rm J} = 1/\sqrt{1+\epsilon^{0}}$}.

\subsection{Axisymmetric steady-state drift solution} \label{sec:nsh}

To demonstrate the scaling of our method with the number of species, we recover the steady-state gas and dust drift solution in a shearing box, as in section 3.5.2 of \cite{BenitezLlambay2019}.
 This test consists of a 1D shearing box along the x direction, with size $L=1$ and with constant frequency $\Omega_0=1$. The shear parameter is set to $q=3/2$. We solve the continuity and $x$-$y$ momentum equations for each fluid. Gas and dust momentum equations include the centrifugal force (${\bf f}_{\rm ce} =\rho 2q\Omega^2_0x \hat{x}$) and the Coriolis force ${\bf f}_c = -2\rho \Omega_0 \hat{z}\times (v_x,v_y)$. In addition, we add an external constant force on the gas fluid that mimics the global pressure gradient in a protoplanetary disk, ${\bf f}_{\rm g}=\chi_0\Omega_0\hat{x}$ with $\chi_0=0.005$.
We add these external forces to the time-explicit hydrodynamic solver in an unsplit fashion \citep[see e.g.][]{Gressel}. 
We apply periodic boundary conditions in $\rho, \rho v_x$. Whereas for $v_y$ we add the relative shear velocity after applying periodic boundary conditions, that is $\rho v_y(x)=\rho v_y(x \mp L) \pm q\Omega_0L\rho$ \citep{Hawley1995}.
The initial condition is  set to the exact equilibrium following Eqs.\,(81-84) of \citep{BenitezLlambay2019}.
We integrate the system until $t=20$ and vary the number of grid points in $x$ from $N_x=32$ to $N_x=1024$ . In all cases, our numerical method conserves the initial equilibrium to machine precision.

\begin{figure}[h]
\includegraphics[scale=0.75]{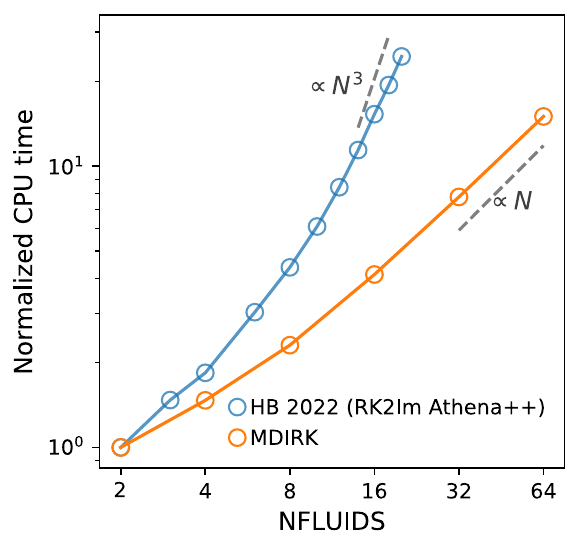}
\caption{Normalized CPU time vs the number of fluids for the test described in Section\,\ref{sec:nsh}. As expected, the complexity of MDIRK is $\mathcal{O}(N)$. For comparison, we also include the results of the multifluid second-order fully implicit method (RK2Im) from \cite{Huang2022} which has complexity $\mathcal{O}(N^3)$.  }
\label{fig:scaling}
\end{figure}

In Figure\,\ref{fig:scaling}, we show the normalized CPU time obtained for this test, where we vary the number of dust species from 1 to 63. As expected, the computational cost is proportional to the number of fluids (complexity $\mathcal{O}(N)$). 
We moreover compare our result with the second-order multifluid method RK2Im of \citep{Huang2022}, which is found to be close to $\mathcal{O}(N^3)$) for a large number of fluids ($N>10$).
It is interesting to note that the complexity of \citet{Huang2022} method can be reduced to $\mathcal{O}(N)$ by utilizing our analytical solution in Appendix\,\ref{sec:asymptotic}.

\bigskip
\bigskip
\bigskip

\section{Discussion and conclusions}
\label{sec:conclusions}
In this work, we have presented MDIRK, a multifluid asymptotically stable numerical method to solve the momentum exchange between gas and dust species. 
We have shown that a hydrodynamical solver based on a Runge-Kutta implicit-explicit (IMEX) algorithm converges to the correct equilibrium solution with second-order accuracy. 
The time-implicit part of the IMEX utilizes a diagonally implicit Runge–Kutta method (DIRK) \citep{Alexander1977}.
The major advantage of a diagonally implicit method for the drag force is that  it admits an analytical solution that reduces the complexity of the problem from $\mathcal{O}(N^3)$ (needed to invert the matrix numerically) to $\mathcal{O}(N)$ (needed to evaluate the solution). 
Our second-order implicit-explicit solver can be easily implemented in state-of-the-art hydrodynamics codes (e.g., PLUTO \citep{Mignone2007}, ATHENA++  \citep{Stone2020}, among others) and it allows for efficient parallelization by fluids through the utilization of of weighted momenta and density \citep[c.f.][]{Krapp2020}. 
Extension to different second-order methods can be also obtained utilizing the analytical solution of Appendix\,\ref{sec:asymptotic}. 

Our results suggest that IMEX Runge-Kutta methods with diagonally implicit drag-force schemes are an interesting alternative for developing high-order multifluid codes.
Examples of third-order and four-order IMEX methods can be found in \citep{ASCHER1997,Pareschi_Russo_2005}. 
Note however that a high-order integration will require monotonic preserving spatial reconstruction techniques with adequate accuracy \citep{Mignone2014,Felker2018}. In addition, we stress that only a subset of high-order Runge-Kutta methods are consistent with a total-variation-diminishing property (TVD) \citep{SHU1988,Gottlieb} and therefore special care must be taken with high-order methods to prevent the presence of spurious oscillation.  

While we focus on 1D tests, our method can be utilized for 2D and 3D with static and adaptive mesh refinement with the appropriate extension of the hydrodynamics \citep{Lebreuilly2019, Huang2022}. 
Also, in this work we did not take into account mass transfer between species, however, future extensions could explore this feature by combining our method with a dust-growth solution based on  discretized dust bins method as described in \cite[]{Lombart2021}. Another important future extension of our method would be the inclusion of energy exchange between dust and gas \citep{Muley2023}, since collisional cooling could play an important role in the thermodynamics of protoplanetary disks. In radiation hydrodynamics,  multi-fluid dynamics can also impact the opacity distribution and observational signatures of accreting, embedded planets \citep{Krapp2022}.
We emphasize that the complexity $\mathcal{O}(N)$ combined with the stability properties will expedite the study of particle-size distribution, especially in problems where hundreds and/or thousands of bins are required to properly model the dust component \citep[see e.g.,][]{Krapp2019,Zhu2021}.

\vspace{0.5cm}

We thank the reviewer for a valuable clear report.
L.\ K.\ and K.~M.~K  acknowledge support  from the Heising-Simons 51 Pegasi b postdoctoral fellowship and TCAN grant 80NSSC19K0639.
P.~B.~L. acknowledges  support from ANID, QUIMAL fund ASTRO21-0039 and FONDECYT project 1231205.
J.G. acknowledge support by ANID, -- Millennium Science Initiative Program -- NCN19\_171, CCJRF2105, FONDECYT Regular grant 1210425 and CATA-BASAL project FB210003.

%%%%%%%%%%%%%%%%%%%%%%%%%%%%%%%%%%%%%%%%%%%%%%%%%%
%\section*{Data Availability}
\vspace{0.5cm}

All codes and data utilized in this work are available in \href{https://bitbucket.org/secondorderriemannsolver/somrs/src/master/}{bitbucket.org/secondorderriemannsolver/somrs}.

\bibliography{biblio}{}
\bibliographystyle{aasjournal}

\appendix

\section{Asymptotic stability}
\label{sec:asymptotic}
In this section, we show that MDIRK is asymptotically stable. As in Section\,\ref{sec:numerical_method_deriv}, we will only focus on the system of momentum equations. 
Moreover, we only consider the implicit part of the method. In this case, MDIRK reduces to a two-stage diagonally implicit method of the form
\begin{align}
	\left( {\bf I} - \gamma \Delta t {\bf M} \right) {\bf k}_1 &= {\bf M} {\bf u}^{n}\,, \label{eq:k1app} \\
	\left( {\bf I} - \gamma \Delta t {\bf M} \right) {\bf k}_2 &= {\bf M} \left({\bf u}^{n} + \beta \Delta t {\bf k}_1\right) \label{eq:k2app} \\
 {\bf u}^{n+1} &= {\bf u}^n +b_1 \Delta t {\bf k}_1 + b_2 \Delta t {\bf k}_2\,,\label{eq:rbs2}
\end{align}
with $\beta = 1-\gamma$, $b_1=1-\gamma$ and $b_2=\gamma$.
The analytical solution of Eq.\,\eqref{eq:rbs2} can be obtained after solving Eqs.\,\eqref{eq:k1app}-\eqref{eq:k2app} for ${\bf k}_1$ and ${\bf k}_2$ utilizing the following summations
\begin{align}
\label{eq:sumABCD}
    \mathcal{A} &= \sum^{N-1}_{k=1} \frac{\alpha_{{\rm d},k} u_{{\rm d},k}}{(1+\gamma \Delta t \alpha_{{\rm d},k}) } \,, \quad
	\mathcal{B}= \sum^{N-1}_{k=1}\frac{ \epsilon_k\alpha_{{\rm d},k}}{(1+\gamma \Delta t \alpha_{{\rm d},k})}\,, \nonumber \\
    \mathcal{C} &= \sum^{N-1}_{k=1} \frac{ \alpha^2_{{\rm d},k} u_{{\rm d},k}}{(1+\gamma \Delta t \alpha_{{\rm d},k})^2 } \,, \quad
	\mathcal{D} = \sum^{N-1}_{k=1}\frac{ \epsilon_k\alpha^2_{{\rm d},k}}{(1+\gamma \Delta t \alpha_{{\rm d},k})^2}\,, 
\end{align}
with

\begin{align}
    \label{eq:Kg}
    k_{1{\rm g}} &= \frac{\mathcal{A}-u_{\rm g}^n\mathcal{B}}{1+\gamma\Delta t \mathcal{B}}  \\
    k_{2{\rm g}} &= k_{1{\rm g}} \left( \frac{1 + (\gamma-\beta)\Delta t \mathcal{B} + \gamma \beta \Delta t^2 \mathcal{D}}{1+\gamma\Delta t \mathcal{B}}\right) + \frac{\beta \Delta t}{(1+\gamma \Delta t \mathcal{B})}\left(u_{\rm g}^n\mathcal{D} - \mathcal{C}\right)
\end{align}
and 
\begin{align}
    \label{eq:Kd}
    k_{1{\rm d},i} &= \frac{\alpha_i}{1+\gamma \Delta t \alpha_i}\left( \epsilon_i u_{\rm g} - u_{{\rm d},i} + \gamma \Delta t \epsilon_i k_{1{\rm g}}  \right)  \\
    k_{2{\rm d},i} &=  \frac{\left(\alpha_i + (\gamma-\beta)\Delta t \alpha^2_i \right)}{\left(1+\gamma \Delta t \alpha_i\right)^2} \left( \epsilon_i u_{\rm g} - u_{{\rm d},i} \right)  + \frac{\alpha_i \beta\Delta t \epsilon_i}{\left(1+\gamma \Delta t \alpha_i\right)^2}  k_{1{\rm g}}  + \frac{\alpha_i \gamma\Delta t \epsilon_i}{\left(1+\gamma \Delta t \alpha_i\right)}  k_{2{\rm g}} \,.
\end{align}
MDIRK does not utilize the full analytical solutions given in Eqs.\,\eqref{eq:Kg}-\eqref{eq:Kd} since the calculation of ${\bf k}_1$ and ${\bf k}_2$ are combined with the explicit updates (see  Section\,\ref{sec:numerical_method}). However, this analytical solution could be implemented in the method of \citep{Huang2022} reducing the from $\mathcal{O}(N^3)$ complexity to $\mathcal{O}(N)$ without shifting to a mixed implicit explicit scheme. Another alternative could be a method based on Strang-Splitting \citep{Strang1968}, however, we anticipate that strang-splitting may not converge to the correct equilibrium (see Appendix\,\ref{sec:ap_strang}).
\subsubsection{Asymptotic stability}
We will show that this solution is asymptotically stable, that is,
\begin{equation}
    \lim_{\Delta t \rightarrow \infty} {\bf u}^{n+1} - {\bf c} = 0\,,
\end{equation}
where ${\bf c}$ is a constant vector. 
In fact, the terminal velocity of all fluids converges to a steady state given by the velocity of the center of mass, $v_{\rm cm}$, defined as
\begin{equation}
    v_{\rm cm} = \frac{ u_{\rm g}+\sum_k u_{{\rm d},k} }{\rho_{\rm g}+\sum_k \rho_{{\rm d},k}} = \frac{v_{\rm g} + \epsilon_k v_{{\rm d},k}}{1+\sum_k \epsilon_k}\,.
\end{equation}
To demonstrate the asymptotic convergence we define  $u_{\rm d} = \sum_k u_{{\rm d},k} $, $\epsilon = \sum_k \epsilon_{k}$, and $X = \left(u_{\rm d}-\epsilon u_{\rm g}  \right)$. With these definitions the summations from Eq.\,\eqref{eq:sumABCD} can be reduced to
\begin{align}
\label{eq:lim_sum}
    \lim_{\Delta t \rightarrow \infty} \Delta t \mathcal{A} &= \gamma^{-1} u_{\rm d} \,, \quad
    \lim_{\Delta t \rightarrow \infty} \Delta t \mathcal{B}= \gamma^{-1} \epsilon \,, \nonumber \\
     \lim_{\Delta t \rightarrow \infty} \Delta t^2 \mathcal{C} &= \gamma^{-2} u_{{\rm d}}\,, \quad \lim_{\Delta t \rightarrow \infty} \Delta t^2\mathcal{D} = \gamma^{-2}\epsilon\,.
\end{align}
Using the limits from Eq.\,\eqref{eq:lim_sum} and the coefficients $\beta=b_1=1-\gamma$ and $b_2=\gamma$ it can be shown that
\begin{align}
\label{eq:Klim}
     \lim_{\Delta t \rightarrow \infty} \Delta t k_{\rm 1 g} &= \gamma^{-1} \frac{X}{1+\epsilon} \nonumber \\
     \lim_{\Delta t \rightarrow \infty} \Delta t k_{\rm 2 g} &=\gamma^{-1} \frac{X}{1+\epsilon}  - \gamma^{-2} \frac{(1-\gamma)}{1+\epsilon} X \nonumber \\
     \lim_{\Delta t \rightarrow \infty} \Delta t k_{{\rm 1 d},k} &= \gamma^{-1}\left(\epsilon_k u_{\rm g} - u_{{\rm d},k}\right) +  \gamma^{-1}\frac{\epsilon_k}{1+\epsilon}X \nonumber \\ 
      \lim_{\Delta t \rightarrow \infty} \Delta t k_{{\rm 2 d},k} &= \gamma^{-2}(2\gamma-1) \left(\epsilon_k u_{\rm g} - u_{{\rm d},k}\right) +  \frac{\epsilon_k }{1+\epsilon}X\left( \gamma^{-1}  - \gamma^{-2} (1-\gamma)\right)\,.
\end{align}

Taking the limit of $\Delta t \rightarrow \infty$ to Eq.\,\eqref{eq:rbs2} using the limits from Eq.\,\eqref{eq:Klim} we obtain
\begin{align}
    \lim_{\Delta t \rightarrow \infty} u^{n+1}_{\rm g} &= u_{\rm g}^{n} + \left( (1-\gamma)\gamma^{-1} + 1 - \gamma^{-1}(1-\gamma) \right) \frac{X}{(1+\epsilon)} \nonumber \\
    &= \frac{\left(u_{\rm d}+u_{\rm g}  \right)}{(1+\epsilon)}\,,
\end{align}
\begin{align}
     \lim_{\Delta t \rightarrow \infty} u^{n+1}_{{\rm d},k} &= u_{{\rm d},k}^{n} \left( 1 - \gamma^{-1}(1-\gamma) - \gamma^{-1}(2\gamma-1)\right) + u_{\rm g} \epsilon_k \gamma^{-1} \left( (1-\gamma) + (2\gamma-1)\right) \nonumber \\
     &+\frac{\epsilon_k }{1+\epsilon} X \left( (1-\gamma)\gamma^{-1} + \gamma\left(\gamma^{-1} - \gamma^{-2}(1-\gamma)\right)\right) \nonumber \\
     &= u_{\rm g}\epsilon_k + \frac{\epsilon_k }{(1+\epsilon)} X  \nonumber \\
     &= \epsilon_k \frac{\left(u_{\rm d}+u_{\rm g}  \right)}{(1+\epsilon)}\,.
\end{align}
In this derivation, it seems that the result does not depend on the value of $\gamma$. However, this is only due to the particular choice of the coefficients $\beta$, $b_1$ and $b_2$. A different set of coefficients will lead to the condition $\gamma=1\pm1/\sqrt{2}$, together with constraint for second-order accuracy. 

\section{Second-order time accuracy}
\label{sec:ap_order}
To demonstrate the second-order accuracy in time of the method described in Section\,\ref{sec:numerical_method_deriv} we consider an equation of the form
\begin{equation}
    \frac{ {\rm d} {\bf u}}{ {\rm d} t} =  \bf{G}( {\bf u}) + {\bf M} {\bf u}
    \label{eq:system}
\end{equation}
where ${\bf G}( {\bf u} )$ is a vector of any external forces i.e. those that do not involve coupling between different fluids, that is ${\bf G}({\bf u})_i = G_i(u_i)$. Forces due to coupling enter only through {\bf M}.
To proceed we define ${\bf u}(t_n) = {\bf u}^n$ and emphasise that the notation ${\bf q}^n$,  $\left( {\bf q}\right)^n$, and $\left[ {\bf q}\right]^n$ indicates the function $q(t)$ evaluated when $t=t_n$, for any arbitrary function $q(t)$. The analytical solution of Eq.\,\eqref{eq:system} is obtained as
\begin{equation}
\label{eq:solsystem}
    {\bf u}(t) = {\bf u}^n + \int^{t}_{t_n} {\bf G}( {\bf u}) \,{\rm d} t + \int^{t}_{t_n} {\bf M} {\bf u} \, {\rm d}t\,.
\end{equation}

We now approximate the solution \eqref{eq:solsystem} to second-order in time. 
First, the conserved variable ${\bf u}(t)$ and the force, ${\bf G}({\bf u})$, can be approximated (to first order in time) as
\begin{align}
    {\bf u}(t) &\simeq {\bf u}^n + (t-t_n)\left( \frac{ {\rm d} {\bf u}}{ {\rm d} t} \right)^n \\ \nonumber
    {\bf G}( {\bf u}) &\simeq  {\bf G}( {\bf u}^n) + ( t - t_n) \left[\nabla_u {\bf G}( {\bf u}) \right]^n \left( \frac{ {\rm d} {\bf u}}{ {\rm d} t} \right)^n 
\end{align}
where $\left[\nabla_u {\bf G}( {\bf u}) \right]^n$ is a diagonal operator with diagonal elements $\left[\nabla_u {\bf G}( {\bf u}) \right]^n_{ii} = \left( {\rm d} G_i(u_i)/{\rm d} u_i\right)^n$. We emphasize again that $G_i$ is only a function of $u_i$.
With this consideration, the solution \eqref{eq:solsystem} can be approximated  as follows
\begin{align}
    {\bf u}(t) &\simeq {\bf u}^n + {\bf G}( {\bf u}^n) (t-t_n) +  \left[\nabla_u {\bf G}( {\bf u}) \right]^n \left( \frac{ {\rm d} {\bf u}}{ {\rm d} t} \right)^n  \frac{\left( t - t_n\right)^2}{2} \nonumber \\ 
    &+ {\bf M} \left( {\bf u}^n (t-t_n) + \left(\frac{ {\rm d} {\bf u}}{ {\rm d} t} \right)^n  \frac{\left( t - t_n\right)^2}{2} \right)\,.
\end{align}
Eq.\,\eqref{eq:system} implies that 
\begin{equation}
    \left( \frac{ {\rm d} {\bf u}}{ {\rm d} t} \right)^n =  {\bf G}( {\bf u}^n) + {\bf M} {\bf u}^n\,,
\end{equation}
we finally obtain
\begin{align}
     {\bf u}(t) &\simeq {\bf u}^n + (t-t_n) {\bf G}( {\bf u}^n) + (t-t_n){\bf M} {\bf u}^n + \frac{(t-t_n)^2}{2} \left( \left[\nabla_u {\bf G}( {\bf u}) \right]^n + {\bf M}  \right) {\bf G}({\bf u}^n) + \frac{(t-t_n)^2}{2} \left( \left[\nabla_u {\bf G}( {\bf u}) \right]^n {\bf M} + {\bf M}^2 \right) {\bf u}^n\,.
     \label{eq:taylor_ana}
\end{align}

\bigskip

We now apply the numerical scheme from Section\,\ref{sec:numerical_method} with the coefficients $\beta=b_1=1-\gamma$ and $b_2=\gamma$. A full update is obtained as follows\footnote{Note that we already shown that this update is equivalent to the scheme described in Section\,\ref{sec:numerical_method}}.
\begin{align}
    \left( {\bf I} - \gamma \Delta t {\bf M}\right) {\bf k}_1 &= {\bf M}\left( {\bf u}^n + \gamma \Delta t {\bf G}({\bf u}^n) \right) \\
    {\bf u}^{*} &= {\bf u}^n + \gamma \Delta t {\bf k}_1 + \Delta t \gamma {\bf G}({\bf u}^n)  \\
    \left( {\bf I} - \gamma \Delta t {\bf M}\right) {\bf k}_2 &= {\bf M}\left( {\bf u}^n + (1-\gamma )\Delta t {\bf k}_1 + \delta \Delta t {\bf G}({\bf u}^n) +  (1-\delta) \Delta t {\bf G}({\bf u}^*) \right)  \nonumber \\
    {\bf u}^{n+1} &= {\bf u}^n + (1-\gamma) \Delta t {\bf k}_1 + \gamma \Delta t {\bf k}_2 + \delta \Delta t {\bf G}({\bf u}^n) + (1-\delta) \Delta t {\bf G}({\bf u}^*)
\end{align}
Defining 
\begin{equation}
    \label{eq:Amatrix}
    {\bf A} = \left( {\bf I} - \gamma \Delta t {\bf M} \right)^{-1} {\bf M}\,,
\end{equation}
the solution follows as
\begin{align}
    {\bf u}^{n+1} &= \left( {\bf I} + \Delta t {\bf A} + \gamma (1-\gamma) \Delta t ^2 {\bf A}^2 \right){\bf u}^{n}  \\ \nonumber
    &+ \left( ( \gamma(1-\gamma) + \gamma \delta ) \Delta t^2 {\bf A} + \gamma^2 (1-\gamma)\Delta t^3 {\bf A}^2 + \delta \Delta t {\bf I}  \right) {\bf G}({\bf u}^n) \\ \nonumber
    &+ \left( \gamma (1-\delta) \Delta t^2 {\bf A} + (1-\delta)\Delta t {\bf I}  \right) {\bf G}({\bf u}^*)\,.
    \label{eq:sol_sys}
\end{align}

To find the second order approximation of the numerical solution\,\eqref{eq:sol_sys} we first note that
\begin{align}
    {\bf A} &\simeq {\bf M} + \Delta t {\bf M}^2    \,,\\
    {\rm G}({\bf u}^*) &\simeq {\bf G}({\bf u}^n) + \gamma \Delta t \left( \left[\nabla_u {\bf G}( {\bf u}) \right]^n {\bf G}({\bf u}^n) + \left[\nabla_u {\bf G}( {\bf u}) \right]^n {\bf M} {\bf u}^n \right)\,,
\end{align}
where we have used that ${\bf k}_1 \simeq \Delta t {\bf M}{\bf u}^n$ to first order.

Replacing these two expansions in Eq.\,\eqref{eq:sol_sys}, and neglecting terms with order higher than $(\Delta t)^2$, we obtain
\begin{align}
    {\bf u}^{n+1} &\simeq \left( {\bf I} + \Delta t ({\bf M} + \Delta t {\bf M}^2) + \gamma (1-\gamma) \Delta t ^2 {\bf M}^2 \right){\bf u}^{n} \nonumber \\
    &+ \left( ( \gamma(1-\gamma) + \gamma \delta ) \Delta t^2 {\bf M} + \delta \Delta t {\bf I}  \right) {\bf G}({\bf u}^n) \\
    &+ \left( \gamma (1-\delta) \Delta t^2 {\bf M} + (1-\delta)\Delta t {\bf I}  \right)  \nonumber \\
    &\times \left[ {\bf G}({\bf u}^n) + \gamma \Delta t \left( \left[\nabla_u {\bf G}( {\bf u}) \right]^n {\bf G}({\bf u}^n) + \left[\nabla_u {\bf G}( {\bf u}) \right]^n {\bf M} {\bf u}^n \right) \right]\,.
\end{align}

which can be cast as
\begin{align}
    {\bf u}^{n+1} &\simeq {\bf u}^n + \Delta t ({\bf M} + \Delta t {\bf G}({\bf u}^n) \\
    &+ \Delta t^2 \left( \gamma (2-\gamma){\bf M}^2 + \gamma (1-\delta) \left[\nabla_u {\bf G}( {\bf u}) \right]^n {\bf M} \right) {\bf u}^n \\
    &+ \Delta t^2\left( \gamma (1-\delta) \left[\nabla_u {\bf G}( {\bf u}) \right]^n 
 + \gamma(2-\gamma){\bf M}\right) {\bf G}({\bf u}^n) \,.
\end{align}
In order to match the Taylor series \eqref{eq:taylor_ana} for $t=t_n+\Delta t$, we must have
\begin{equation}
    \gamma(2-\gamma) = \frac{1}{2}\,, \quad \quad \quad \gamma(1-\delta) = \frac{1}{2}
\end{equation}
These two conditions give
\begin{equation}
    \gamma = 1 \pm \frac{1}{\sqrt{2}}\,, \quad {\rm and,} \quad \delta = 1 - \frac{1}{2 \gamma}\,,
\end{equation}
which is the solution from \cite{ASCHER1997}.

\section{Convergence to the correct equilibrium}
\label{sec:ap_conv}

To study the convergence properties of the numerical method we consider the case of Eq.\,\eqref{eq:system} with constant external forces ${\bf g}({\bf u}) = {\bf g}_0$.  In this case, the second-order Runge-Kutta IMEX method described in Section\,\ref{sec:numerical_method_deriv} reduces to 
\begin{align}
\label{eq:sol_constant}
    {\bf u}^{n+1} &= \left( {\bf I} + \Delta t {\bf A} + \gamma (1-\gamma) \Delta t ^2 {\bf A}^2 \right){\bf u}^{n} \nonumber \\ 
    &+ \left( \gamma(2-\gamma)  \Delta t^2 {\bf A} + \gamma^2 (1-\gamma)\Delta t^3 {\bf A}^2 + \Delta t {\bf I}\right) {\bf G}_0\,.
\end{align}
Since ${\bf A}$ and ${\bf A}^2$ are diagonalizable under the same transformation (see Section\,\ref{ap:Adiag}), we apply the transformation matrix ${\bf P}^{-1}$ to Eq.\,\eqref{eq:sol_constant} and obtain
\begin{align}
\label{eq:conv}
    \hat{u}^{n+1}_k &= \hat{u}^{n}_k \left( 1 + \Delta t \frac{\lambda_k}{1-\gamma \Delta t \lambda_k} + \gamma (1-\gamma) \Delta t^2 \left( \frac{\lambda_k}{1-\gamma \Delta t \lambda_k} \right)^2\right) \nonumber \\
    &+ \hat{G}_k \left[ \gamma(2-\gamma) \Delta t^2 \frac{\lambda_k}{1-\gamma\Delta t \lambda_k} + \gamma^2 (1-\gamma) \Delta t ^3 \left( \frac{\lambda_k}{1-\gamma\Delta t \lambda_k}\right)^2 + \Delta t \right]\,,
\end{align}
where $\hat{ {\bf u}} = {\bf P}^{-1}{\bf u}$ and $\hat{u}_k$ corresponds to the entry of $\hat{{\bf u}}$ associated to the eigenvalue $\lambda_k$ of the matrix ${\bf M}$ (equivalently for $\hat{G}_k$).
As mentioned in Section\,\ref{sec:numerical_method_intro}, all eigenvalues of the matrix ${\bf M}$ are negative, except for a unique eigenvalue $\lambda_j=0$. 
We define $x = |\lambda_k| \Delta t$ (with $|\lambda_k| = -\lambda_k$) and write Eq.\,\eqref{eq:conv} as
\begin{align}
    \hat{u}^{n+1}_k = \hat{u}^n_k\left(\frac{ x(2\gamma-1) +1 }{(1+\gamma x)^2} \right) +   \frac{ \hat{G}_k}{|\lambda_k|} \frac{ x(\gamma^2 x +1) }{(1+\gamma x)^2}\,.
\end{align}
For the case of the eigenvalue $\lambda_j=0$, $\hat{u}^{n+1}_j = \hat{u}^{n}_j + \Delta t \hat{G}_j$, which corresponds to a uniformly accelerated center of mass of the system. For the case with $\lambda_k \neq 0$, the limit of $x\rightarrow\infty$ gives
\begin{equation}
    \lim_{x\rightarrow \infty} \hat{u}^{n+1}_k = \lim_{x\rightarrow \infty} \left(\hat{u}^{n}_k \frac{2\gamma-1}{\gamma^2 x} + \frac{ \hat{G}_k}{|\lambda_k|}\left(1 + \frac{1-2\gamma}{\gamma^2 x}\right) \right) = \frac{ \hat{G}_k}{|\lambda_k|}\,.
\end{equation}
This limit is identical to that obtained after applying the transformation ${\bf P}^{-1}$ to the analytical solution\,\eqref{eq:solsystem} with a constant external force ${\bf G}={\bf G}_0$ (see Eq.\,\eqref{eq:sol_damping_g_lim}). Thus, the solution corresponds to a system with constant velocities relative to the center of mass.
Note as well that in cases with no external acceleration $ \lim_{x\rightarrow \infty} \hat{u}^{n+1}_k \propto \lim_{x\rightarrow \infty} 1/x$,  which explains the asymptotic convergence of the error as $e \sim 1/\Delta t$ obtained in Section\,\ref{sec:athena_test}.

\subsection{Strang-splitting}
\label{sec:ap_strang}
We now show that a strang-splitting method does not converge to the correct equilibrium solution. Consider a strang-splitting method where the drag force is integrated on $\Delta t/2$ before and after a full integration of the hydrodynamics.
To study the convergence of the strang-splitting method we assume ${\bf u}^n = 0$  since the initial condition should not affect the convergence to the correct equilibrium.
Thus, we skip the first implicit update on the half step. The solution of the strang-splitting method is obtained as follows
\begin{align}
\label{eq:sol_constant}
    {\bf u}^{n+1} &= \left( {\bf I} + \frac{1}{2}\Delta t {\bf A} + \gamma (1-\gamma) \frac{1}{4}\Delta t ^2 {\bf A}^2 \right) \Delta t{\bf G}_0\,.
\end{align}
Following the diagonalization steps and normalization described for Eq.\,\eqref{eq:conv} we obtain
\begin{equation}
     \hat{u}_k^{n+1} = \frac{\hat{G}_k}{|\lambda_k|} x \left( 1 - \frac{x/2 (1-\gamma)}{1+\gamma x/2} - \frac{x/2 \gamma}{(1+\gamma x/2)^2} \left( 1+ x/2 (2 \gamma -1) \right) \right)
\end{equation}
for some eigenvalue $\lambda_k \neq 0$. 
Now we take the limit of $x\rightarrow\infty$ and obtain
\begin{equation}
    \lim_{x\rightarrow\infty} \hat{u}_k^{n+1} = \left( \frac{\hat{G}_k}{|\lambda_k|}  \right) \lim_{x\rightarrow\infty} x \left( 1 - \frac{x/2 (1-\gamma)}{1+\gamma x/2} - \frac{x/2 \gamma}{(1+\gamma x/2)^2} \left( 1+ x/2 (2 \gamma -1) \right) \right) = \left( \frac{\hat{G}_k}{|\lambda_k|}  \right) \frac{4\gamma -2}{\gamma^2}.
\end{equation}

The convergence to the correct equilibrium implies $4\gamma-2/\gamma^2 = 1$ which gives $\gamma=2\pm \sqrt{2}$. However, this value is not consistent with the second-order accuracy requirement discussed in Section\,\ref{sec:numerical_method}. Therefore, the strang-splitting fashion will not converge to the correct steady-state solution.

\subsection{ The matrix ${\bf A}$ is diagonalizable}
\label{ap:Adiag}

First note that ${\bf M}$ is diagonalizable since it is similar to a real symmetric matrix as shown in \cite{BenitezLlambay2019}. Now, we define ${\bf P}$ as the transformation matrix such that ${\bf P}^{-1}{\bf M}{\bf P} = {\bf \Lambda}$, with ${\bf \Lambda}$ a diagonal matrix. 
The transformation defined by ${\bf P}$ also diagonalize ${\bf I} - \gamma \Delta t {\bf M}$, together with the inverse $\left( {\bf I} - \gamma \Delta t {\bf M}\right)^{-1}$. Therefore, 
\begin{align}
    {\bf P}^{-1}{\bf A}{\bf P} &= {\bf P}^{-1} \left( \left( {\bf I} - \gamma \Delta t {\bf M}\right)^{-1}{\bf M}\right){\bf P} \nonumber \\
    &=  \left( {\bf P}^{-1} \left( {\bf I} - \gamma \Delta t {\bf M}\right)^{-1} {\bf P} \right) \left( {\bf P}^{-1}{\bf M}{\bf P} \right) = {\bf D} {\bf \Lambda} = {\bf A}_{\rm D}
\end{align}
where ${\bf D}$ and  ${\bf A}_{\rm D}$  are diagonal matrices. 

\section{On the value of $\gamma$}
\label{ap:gama}

We stress that $\gamma=1-1/\sqrt{2}$ (instead of $\gamma=1+1/\sqrt{2}$)  will provide the most accurate solution since the truncation error is minimized for this value. However, depending on the problem at hand,  the value of $\gamma=1-1/\sqrt{2}$ may not be consistent with a TVD Runge-Kutta in the form of \citep{SHU1988}. 
For example, the explicit part of the solver in Section\,\ref{sec:numerical_method} can be written as
\begin{align}
{\bf u}^{(0)} &= {\bf u}^n \nonumber \\    
{\bf u}^{(1)} &= \alpha_{10} {\bf u}^0 + \beta_{10}\Delta t \mathcal{H}({\bf u}^0) \nonumber \\  
{\bf u}^{(2)} &= \alpha_{20} {\bf u}^{(0)} + \alpha_{21} {\bf u}^{(1)} + \beta_{20}\Delta t\mathcal{H}({\bf u}^{(0)}) + \beta_{21}\Delta t \mathcal{H}({\bf u}^{(1)}) \nonumber \\
{\bf u}^{n+1} &= {\bf u}^{(2)}
\end{align}
with $\alpha_{10}=1$, $\alpha_{20}=1$, $\alpha_{21}=0$, $\beta_{10}=\gamma$,  $\beta_{20}=\delta$, $\beta_{21}=1-\delta$. 
These values are consistent with TVD second-order Runge-Kutta 
provided  $\gamma=1+1/\sqrt{2}$, since all the coefficients of the explicit update are non-negative \citep{Gottlieb}. However, when $\gamma=1-1/\sqrt{2}$ the coefficient $\beta_{20}=\delta <0$.

There is yet another concern about the value $\gamma = 1-1/\sqrt{2}$.
To showcase the problem, we consider a reduced system where only drag forces are taken into account. In this case, the numerical solution of the system of equations\,\eqref{eq:mom_gas}-\eqref{eq:mom_dust} reduces to the form
\begin{equation}
\label{eq:unique}
 {\bf u}^{n+1} =  \left( {\bf I} + \Delta t {\bf A} + \gamma(1-\gamma) \Delta t ^2 {\bf A}^2  \right) {\bf u}^{n} \equiv  {\bf T}(\Delta t, \gamma)  {\bf u}^{n} \,
\end{equation}
where ${\bf A}$ is defined in Eq.\,\eqref{eq:Amatrix}.
If for a given time step the $\rm{det} \left( {\bf T}(\Delta t, \gamma) \right) = 0$,  the solution ${\bf u}^{n+1}$ may deviate from the correct instantaneous equilibrium. 
%
%In the following, we will show that there exists a finite $\Delta t >0$ such $\rm{det} \left( {\bf T}(\Delta t, \gamma) \right) = 0$ for  $\gamma = 1 - 1/\sqrt{2}$ , whereas $\rm{det} \left( {\bf T}(\Delta t, \gamma) \right) \neq 0$ for all $\Delta t > 0$ if $\gamma = 1+1/\sqrt{2}$.
%
First, note that the eigenvalues, $a_k$, of ${\bf A} = ({\bf I} - \gamma \Delta t {\bf M})^{-1} {\bf M}$ are
\begin{equation}
    a_k = \frac{\lambda_k}{1-\gamma \Delta t \lambda_k}\,,
\end{equation}
with $\lambda_k$ eigenvalue of ${\bf M}$. Now, we define $\mu_k$ the eigenvalues of ${\bf T}$ and note that 
%
%\begin{equation}
%    \mu_k = 1 + \Delta t a_k + \gamma (1-\gamma) %\Delta t^2 a^2_k
%\end{equation}
%
%
\begin{equation}
    \mu_k = 1 + \Delta t \frac{\lambda_k}{1-\gamma \Delta t \lambda_k} + \gamma (1-\gamma) \Delta t^2 \left( \frac{\lambda_k}{1-\gamma \Delta t \lambda_k} \right)^2\,.
\end{equation}
We introduce the variable $x = \Delta t |\lambda_k|$ as in Section\,\ref{sec:ap_conv} with $|\lambda_k| = -\lambda_k$. Thus, 
\begin{equation}
    \mu_k(x) = 1 - \frac{x}{1+\gamma x} + \gamma (1-\gamma)  \frac{x^2}{(1+\gamma x)^2} \,.
\end{equation}
which can be reduced to
\begin{equation}
    \mu_k(x) = \frac{ x(2\gamma-1) +1 }{(1+\gamma x)^2}\,.
\end{equation}
We conclude that $\mu_k(x) = 0$ for $x = 1/(1-2\gamma)$ and therefore 
\begin{equation}
\label{eq:det}
    {\rm det}( {\bf T} ) = \prod_k \mu_k = 0 \iff  x = 1/(1-2\gamma) \,.
\end{equation}
In terms of $\Delta t$, the condition \eqref{eq:det} implies
\begin{equation}
     {\rm det}( {\bf T} ) = 0 \iff \Delta t = \frac{|\lambda_k|^{-1}}{1-2\gamma}\,,
\end{equation}
for any $k \in [0,N-1]$.
In the case of $\gamma = 1 - 1/\sqrt{2}$, $1-2\gamma >0$ and therefore there is $\Delta t >0$ such that $ {\rm det}( {\bf T} ) = 0 $. In the other hand,  if $\gamma = 1 + 1/\sqrt{2}$, $1-2\gamma <0$ and therefore ${\rm det} ({\bf T}) \neq 0 $ for all $\Delta t>0$.
The same result also holds for the second-order Runge-Kutta of \cite{Pareschi_Russo_2005}.
Finally, the same analysis can be applied to a case with constant external force, however, in that case, the condition that could affect the solution corresponds to $x=-1/\gamma^2$, which is never satisfied for $\Delta t >0$.

\section{Flux calculation}\label{sec:hydro}

To calculate the mass and momentum flux ($\mathcal{L}$ operator neglecting external forces)  we consider a 1D Cartesian evenly spaced mesh along the $x$-direction. For each fluid, the operator $\mathcal{L}$ in Eq.\,\eqref{eq:scheme_full} returns the divergence of the flux as follows
\begin{equation}
    \mathcal{L}({\bf U}_i^n) = - \frac{1}{\Delta x} \left({\bf F}_{i+1/2} ({\bf{U}}^n)- {\bf F}_{i-1/2}({\bf{U}}^n)\right) \,,
\end{equation}
where the index $i$ denotes the spatial position in the discretized mesh and $i \pm 1/2$ indicates the  position of the interfaces where fluxes are calculated.
${\bf U}=(\rho, \rho v)$ correspond to the vector of conserved variables. 
The flux, ${\bf F}_{i+1/2}$, is obtained by solving the Riemann problem with initial condition
\begin{equation}
     {\bf U}(x,0) = \left\{
     \begin{array}{ccc}
       {\bf U}_{{\rm L},i+1/2} & \text{if} & x<x_{i+1/2} \,, \\
       {\bf U}_{{\rm R},i+1/2} & \text{if} & x>x_{i+1/2} \,, \\
     \end{array}
   \right.
\end{equation}
where ${\bf U}_{{\rm L},i+1/2}$ and ${\bf U}_{{\rm R},i+1/2}$ correspond to the left and right states with respect to the interface at $x=x_{i+1/2}$.
In this work, we use a piece-wise linear reconstruction with a Van-Leer slope limiter \citep{VanLeer} on the primitive variables ${\bf V}=(\rho, v)$. We approximate the left and right states as follows
\begin{equation}
    {\bf V}_{{\rm L},i+1/2} = {\bf V}_{i} + \frac{1}{2}\Delta_x {\bf V}_{i} \,, \quad {\bf V}_{{\rm R},i+1/2} = {\bf V}_{i+1} - \frac{1}{2}\Delta_x {\bf V}_{i+1} \,,
\end{equation}
with 
\begin{equation}
\Delta_x {\bf V}_{i} =
\frac{{\rm max} \left\lbrace({\bf V}_{i+1}-{\bf V}_{i})({\bf V}_{i}-{\bf V}_{i-1}),0 \right\rbrace}{{\bf V}_{i+1}-{\bf V}_{i-1}} \,. 
\end{equation}
After the reconstruction of the left and right states calculate the fluxes utilizing the approximate solution from Harten, Lax and van Leer (HLL) \citep{Harten},
\begin{equation}
\label{eq:flux_hll}
    {\bf F}_{i+1/2} = \left\{
    \begin{array}{ccc}
    {\bf F}_{{\rm L}} & \textrm{if  } & 0 < S_{\rm L} \,,\\[-0.5em]
    \\ \frac{\displaystyle S_{\rm R} {\bf F}_{\rm L} - S_{\rm L} {\bf F}_{\rm R} + S_{\rm R}S_{\rm L} ( {\bf U}_{\rm R} - {\bf U}_{\rm L}  ) }{\displaystyle S_{\rm R}-S_{\rm L}} & \textrm{if} & S_{\rm L} \leq 0 \leq S_{\rm R} \,, \\[-0.5em]
    \\ {\bf F}_{{\rm R}} & \textrm{if} & 0 > S_{\rm R} \,, \\
    \end{array}
    \right.
\end{equation}
where ${\bf F}_{{\rm L}}$ and ${\bf F}_{{\rm R}}$
are fluxes computed with ${\bf V}_{{\rm L},i+1/2}$ and ${\bf V}_{{\rm R},i+1/2}$, respectively. 
Thus, in terms of primitive variables, we define the vector fluxes as ${\bf F}_{\rm g} = (\rho_{\rm g} {v}_{\rm g}, \rho_{\rm g}{v}_{\rm g}^{2} + P)$, ${\bf F}_{\rm d} = (\rho_{\rm d} {v}_{\rm d}, \rho_{\rm d}{v}_{\rm d}^{2})$, for gas and dust, respectively. For the gas species, the speeds $S_{\rm R}$ and $S_{\rm L}$ are estimated as
\begin{equation}
    S_{\rm L} = {\rm min}\lbrace v_{\rm L} - c_{{\rm s}}, v_{\rm R} - c_{{\rm s}}\rbrace \,, \quad S_{\rm R} = {\rm max}\lbrace v_{\rm L} + c_{{\rm s}}, v_{\rm R} + c_{{\rm s}} \rbrace \,,
\end{equation}
where $v_{\rm L}$ and $v_{\rm R}$ are obtained after transforming the conservative left and right states into primitive variables.
For pressureless fluids (dust), the sound speed is set to zero, $S_{\rm L} = - S_{\rm R}$, and $S_{\rm R} = {\rm max} \lbrace |v_{{\rm L}}|,|v_{{\rm R}}| \rbrace$. Therefore, for the dust species the flux in Equation \eqref{eq:flux_hll} reduces to the Rusanov flux \citep{RUSANOV1970}. 
We have also implemented the Riemann solver for the dust described in \citep{Huang2022}. We utilize this method in the shock test and steady-state gas and dust drift calculation.
While in this work we adopted the (perhaps) simpler flux estimation, there are different alternatives for the gas \citep[see e.g.,][]{Toro_2009}, and the dust as a pressureless fluid \cite{Leveque2004,Paardekooper2006}.

\section{Gravitational potential}
\label{sec:grav_pot}

The gravitational potential is included as a source term in Eqs.\,\eqref{eq:mom_gas}-\eqref{eq:mom_dust}. 
To incorporate the potential into our numerical method  we solve the discretized form of the Poisson equation using a sparse linear solver with periodic boundary conditions,
\begin{equation}
    \frac{\Phi_{i+1}-2\Phi_{i}-\Phi_{i-1}}{\Delta x^{2}} = 4\pi G \rho_{i}^{n} \,,
\end{equation}
and obtain $\Phi_{i}$ and include the potential into the $\mathcal{L}$ operator as
\begin{align}
    \mathcal{L}({\bf U}^n_i) &=  - \frac{\Delta t}{\Delta x} \left( \left({\bf F}_{i+1/2} ({\bf U}^n)- {\bf F}_{i-1/2}({\bf U}^n)\right) - \frac{\rho_i^{n}}{2}\left(\Phi_{i+1}-\Phi_{i-1}\right) \right)\,.
\end{align}

\section{Matrix ${\bf M}_{\rm BKP19}$}
\label{app:Mbkp}

\begin{align}
\label{eq:Mmatrix}
{\bf M}_{\rm BKP19} &=  
\begin{pmatrix}
 \sum^{N-1}_{k = 1} \epsilon_k \alpha_{{\rm d},k} &   -\epsilon_1\alpha_{{\rm d},1} &  -\epsilon_2 \alpha_{{\rm d},2} & \dotsc &  -\epsilon_{N-1}\alpha_{{\rm d},N-1} \\
 -\alpha_{{\rm d},1} &  \alpha_{{\rm d},1}  & 0 & \dotsc & 0 \\
-\alpha_{{\rm d},2} & 0 & \alpha_{{\rm d},2} & \dotsc & 0 \\
\dotsc  & \dotsc & \dotsc & \dotsc & \dotsc \\
-\alpha_{{\rm d},N-1} & 0 & 0 & \dotsc & \alpha_{{\rm d},N-1} \\
\end{pmatrix} \,.
\end{align}

\end{document}